\documentclass[aps,prd,english,preprintnumbers,nofootinbib,floatfix,twocolumn,10pt]{revtex4-1}

\usepackage{amsfonts,amsmath,amssymb}
\usepackage{graphicx,epsfig}
\usepackage{float}
\usepackage{subfig}
\usepackage{subfloat}
\usepackage[utf8]{inputenc}
\usepackage{hyperref}
\usepackage{babel}
\usepackage[justification=justified]{caption}

\usepackage[font=footnotesize,labelfont=rm,justification=raggedright]{caption}

\newcommand\be{\begin{equation}}
\newcommand\ee{\end{equation}}
\newcommand\bea{\begin{eqnarray}}
\newcommand\eea{\end{eqnarray}}

\begin{document}

\def\rhoo{\rho_{_0}\!} 
\def\rhooo{\rho_{_{0,0}}\!} 

\begin{flushright}
\phantom{
{\tt arXiv:2006.$\_\_\_\_$}
}
\end{flushright}

{\flushleft\vskip-1.4cm\vbox{\includegraphics[width=1.15in]{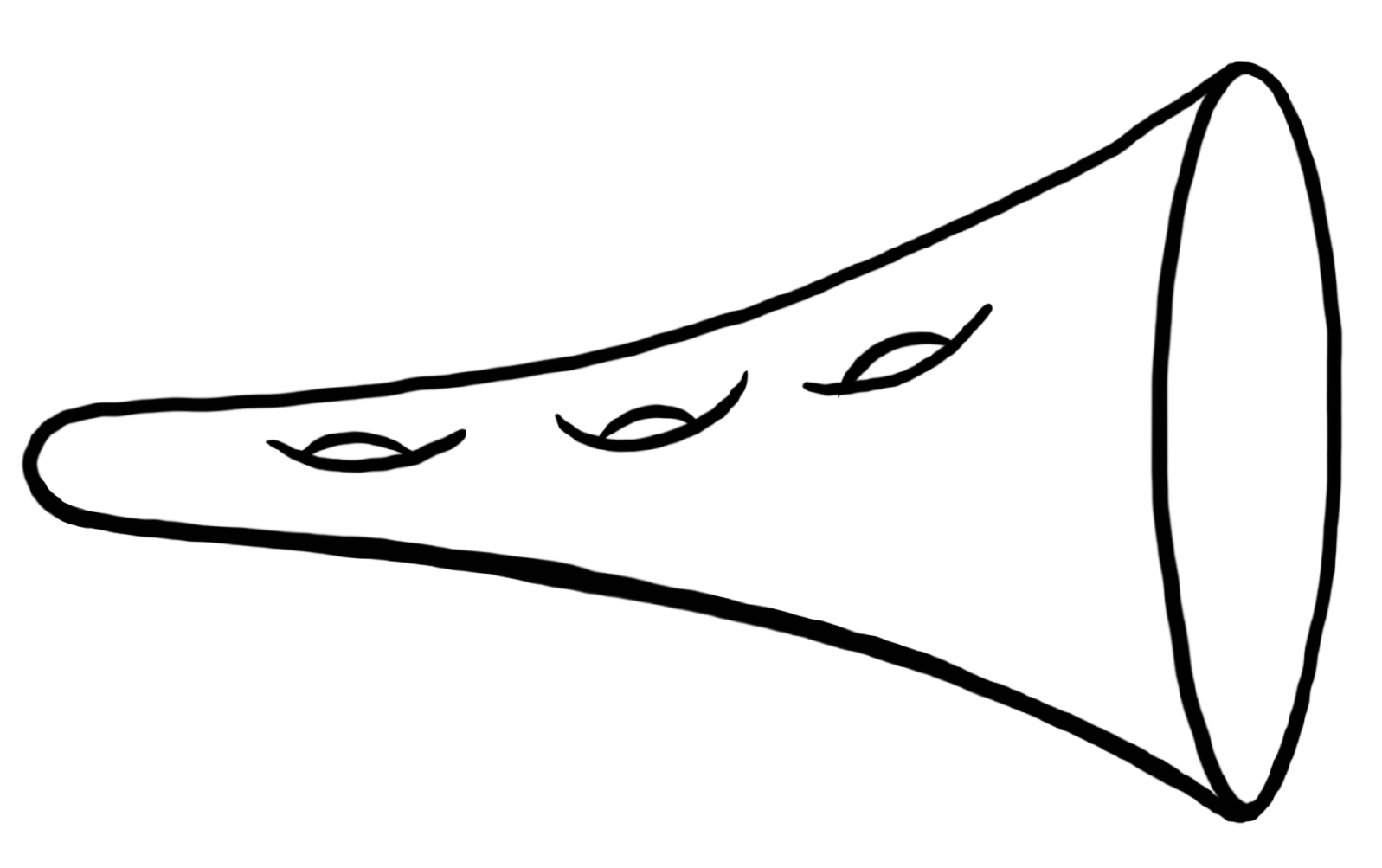}}}

\title{Low Energy Thermodynamics of JT Gravity and Supergravity}
\author{Clifford V. Johnson}
\email{johnson1@usc.edu}
\affiliation{Department of Physics and Astronomy\\ University of
Southern California \\
 Los Angeles, CA 90089-0484, U.S.A.}


\begin{abstract}
Aspects of the low energy physics of certain Jackiw--Teitelboim  gravity and supergravity theories  are explored, using their recently presented non--perturbative description in terms of   minimal string models.  This regime necessarily involves  non--perturbative phenomena, and the inclusion of wormhole geometries connecting multiple copies of the nearly AdS$_2$ boundary in the computation of ensemble averages of key quantities. A new ``replica--scaling'' limit is considered,  combining the replica method and double scaling with the low energy limit.  Using it, the leading free energy, entropy, and specific heat are explored for various examples.  Two models of particular note are the  JT supergravity theory defined as a (1,2) Altland--Zirnbauer matrix ensemble  by Stanford and Witten, and the Saad--Shenker--Stanford matrix model of ordinary JT gravity  (non--perturbatively improved at low energy). The full models have a finite non--vanishing  spectral density at zero energy.  The replica--scaling construction suggests for them  a low temperature entropy and specific heat that are linear in temperature. 

\end{abstract}


\maketitle

\section{Introduction}
The Jackiw--Teitelbiom~\cite{Jackiw:1984je,Teitelboim:1983ux}  (JT) gravity and supergravity systems are  particularly interesting models of two dimensional gravity. They  are instructive examples of low dimensional AdS/CFT  holography,  being in some sense dual (at least for low energy~\cite{Maldacena:2016hyu,Jensen:2016pah,Maldacena:2016upp,Engelsoy:2016xyb}) to models of Sachdev--Ye--Kitaev (SYK) type~\cite{Sachdev:1992fk,Kitaev:talks}. This  makes them a clean laboratory for studying aspects of quantum chaotic dynamics in black holes~\cite{Maldacena:2015waa} and even systems of terrestrial experimental interest.  They also arise naturally from dimensional reduction of various higher dimensional black hole and black brane systems~\cite{Achucarro:1993fd,Fabbri:2000xh,Cvetic:2016eiv,Nayak:2018qej,Kolekar:2018sba,Castro:2018ffi,Ghosh:2019rcj}. Furthermore, JT gravity and supergravity are interesting as models of quantum gravity, dynamically summing over all spacetime geometries, topologies and beyond. As such they have a description in terms of double--scaled random matrix ensembles of various types~\cite{Saad:2019lba,Stanford:2019vob}, and thereby supply another point of intersection between quantum chaos and random matrices~\cite{Guhr:1997ve,M_ller_2004}. Finally, since there are black hole solutions of these solvable models of gravity, by including matter interactions, the JT/SYK setting becomes a powerful laboratory for understanding aspects of the black hole information paradox (early work includes refs.~\cite{Almheiri:2014cka,Krishnan:2017txw,Kourkoulou:2017zaj}).

The primary goal of this paper will be to explore and clarify aspects of the novel low energy properties of certain models  of JT supergravity (although some of the lessons have implications for JT gravity which will be spelled out). This will necessarily involve probing  non--perturbative physics.  There has been some recent success in defining and studying non--perturbative properties~\cite{Johnson:2019eik,Johnson:2020heh,Johnson:2020exp} of JT gravity and supergravity by using a decomposition in terms of   minimal string  models. This is the formalism that will be used for the low energy exploration. In doing so, many new aspects of the JT connection to minimal string physics will be elucidated here for the first time.

A major motivation for exploring the low energy regime arose from observations made in refs.~\cite{Stanford:2019vob,Johnson:2019eik,Johnson:2020heh,Johnson:2020exp} about the full non--perturbative spectral density $\rho(E)$ of various models. Non--perturbative quantum effects can modify the classical spectral density quite dramatically at low energy.  It is already notable when it cancels  a classically divergent spectral density to zero at $E{=}0$ (as happens for the (2,2) JT supergravity model---the notation refers to the $(\boldsymbol{\alpha},\boldsymbol{\beta})$ Altland--Zirnbauer classification scheme~\cite{Altland:1997zz}). However, a {\it finite} non--zero value $\rho(0)$ for the spectral density at $E{=}0$ is particularly interesting, and it is present in the (1,2) JT supergravity, and (in a sense\footnote{\label{foot:low-E}Ordinary JT gravity actually has non--perturbative difficulties  connected to a tail of $E{<}0$ contributions to the spectrum,  but that part of the spectrum can be excised for the purposes of studying the $\rho(0){\neq}0$ feature. This turns out to be justified {\it post--hoc} by noting that a model that naturally modifies the low energy sector in this way exhibits the $\rho(0){\neq}0$ feature~\cite{Johnson:2019eik,Johnson:2020exp}.})  also ordinary JT gravity.   These are also both models without time--reversal symmetry (unlike the two other models, (0,2) and (2,2), for which explicit non--perturbative features have been uncovered) and so it is particularly interesting that they exhibit this feature.

From a condensed matter physics perspective, a quantum mechanical model with a finite spectral density at $E{=}0$ is an interesting  phase with a long and rich history, both experimentally and theoretically. Disordered and ``glassy'' phases such as those that appear in certain magnetic/spin systems can have this feature, as well as other systems with frustrated ground states. Being able to have a quantum gravitational description of such behaviour is novel, and may well be extremely useful. 

\begin{figure}[h]
\centering
\includegraphics[width=0.4\textwidth]{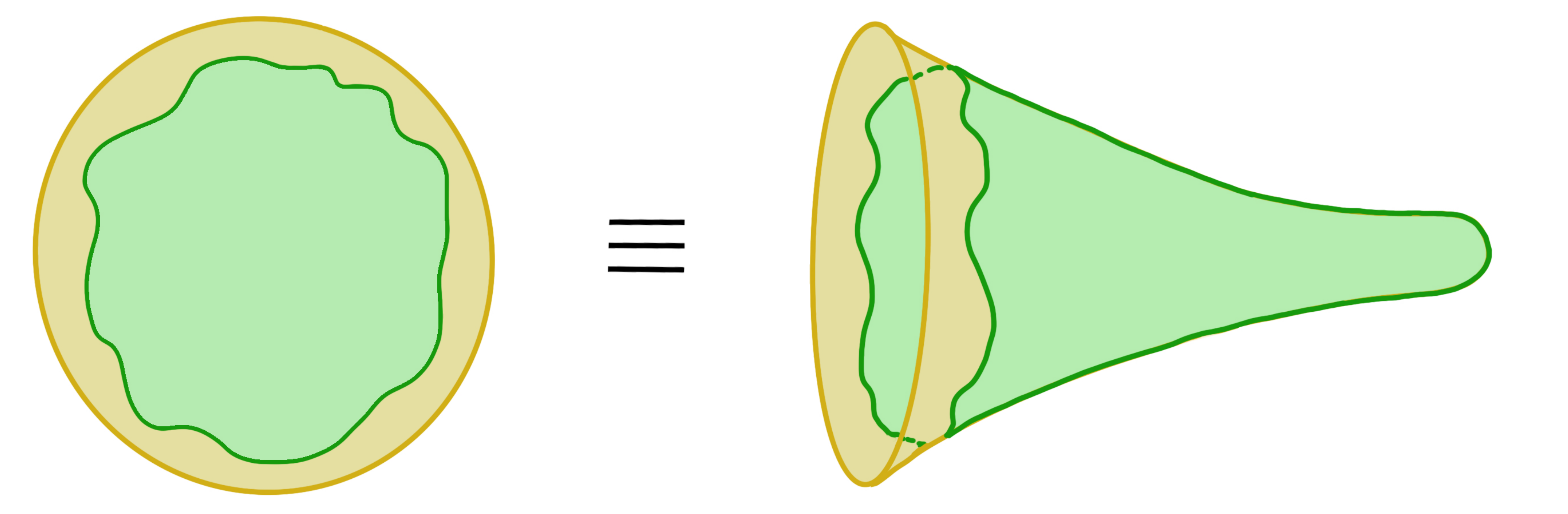}
\caption{\label{fig:disc-diagram} The ``nearly AdS$_2$'' geometry, presented in two equivalent ways.}
\end{figure}

There is also a wider issue of how to describe the low temperature  dynamics of JT gravity and supergravity, even in models when the novelty of  $\rho(0){\neq}0$  is not present.  For example (defining the inverse temperature  $\beta{=}1/T$, and $\hbar{=}e^{-S_0}$ where the constant $S_0$ is the $T{=}0$ entropy), the partition function of JT gravity that arises from the leading (disc order) dynamics is:
\be
\label{eq:JT-partition}
\langle Z(\beta)\rangle_0 = \frac{e^{\frac{\pi^2}{\beta}}}{4\hbar \sqrt{\pi}\beta^{3/2}}\ .\qquad {\rm (JT)}
\ee
As a reminder, the bulk of the theory is AdS$_2$, with all the dynamics residing on the boundary of finite length $\beta$ (so it is ``nearly'' AdS$_2$) whose shape is allowed to fluctuate, according to  a Schwarzian action (see figure~\ref{fig:disc-diagram}).
 The expression above  should be read as an average, or expectation value $\langle \sum_i \exp(-\beta E_i)\rangle$. So when $\beta$ is small, the physics may be safely thought of as dominated by a high energy classical configuration, and the free energy~$F$ can be consistently taken to be the logarithm of  expression~(\ref{eq:JT-partition}) (times $-\beta^{-1}$), while the energy $U$ is the $({-})\beta$--derivative of it.  This yields an entropy, $S{=}\beta(U{-}F)$ which makes sense at high temperature: 
\be
\label{eq:high-T-S}
S=S_0 + \frac{2\pi^2}{\beta} - \frac32\ln\beta + {\rm const.}\ ,
\ee
but which cannot be extended to low temperatures, as signalled by a diverging (and worse, negative)  entropy\footnote{The Author thanks Robie Hennigar, Felipe Rosso, and Andy Svesko for discussions about this in February 2020.}. The story for JT supergravity is similar. (There is a factor of $-\frac12$ in front of the logarithm instead.) 

Crucially, a quick calculation shows that this problem is {\it not} solved by using the full non--perturbative partition function supplied by the studies of ref.~\cite{Johnson:2020exp}. This is all because at low temperature the logarithm of the average partition function is not the right object to study, as is well known from studies in condensed matter physics. Instead, the average $\langle \ln Z(\beta)\rangle$ is the sensible quantity, and it is from this that the energy and entropy should be computed. (This is often referred to as the distinction between the ``annealed'' and the ``quenched'' averaged quantities.)

This latter average is more subtle  to compute in general. In fields such as condensed matter, the replica method is often used, going as far back as the classic work of ref.~\cite{Edwards_1975} (see {\it e.g.,} ref.~\cite{Denef:2011ee} for a review). It is natural to ask if such a method might have a gravitational analogue in JT gravity, and what results may come from it. There has been a lot of discussion of replica methods of different kinds in the JT context in the recent literature\footnote{\label{foot:grav-refs}This began with refs.~\cite{Penington:2019kki,Almheiri:2019qdq}, following on work on entropy and the Page curve in refs.~\cite{Penington:2019npb,Almheiri:2019psf}. See also refs.~\cite{Bousso:2019ykv,Gautason:2020tmk,Hartman:2020swn,Giddings:2020yes,Marolf:2020xie} for closely adjacent work.}, and moreover the seemingly inevitable presence of wormholes in the computations has been interpreted 
as evidence that the gravity path integral should be interpreted as an ensemble average of the dual theory (located on the boundary), which is borne out by the matrix model equivalence~\cite{Saad:2019lba}.  (See also refs.~\cite{Garcia-Garcia:2016mno,Cotler:2016fpe}. For some recent work on an extension of this idea to 2+1 dimensions, see refs.~\cite{Afkhami-Jeddi:2020ezh,Maloney:2020nni}.)

Very recently, ref.~\cite{Engelhardt:2020qpv}  directly posed the question of computing the JT gravity free energy with  a path integral replica computation that includes wormholes, also motivated (at least in large part) by trying to understand the low energy sector. By directly using the matrix model description\footnote{The Author thanks Juan Maldacena, Herman Verlinde, and  Edward Witten for   each (independently) recently asking him the question to as to whether such a computation was possible.}, the methods of this paper are a complement to that approach. The two key advantages of the double--scaled matrix model methods will be, on the one hand, ready access to non--perturbative contributions, and on the other, the efficient handling of (generalized) wormhole contributions.

This paper's results help to emphasize that a non--perturbative treatment is {\it essential} if all of the elements the replica method needs are to be included.  Since non--perturbative physics  can have explicit   dramatic effects on the low energy physics, it can significantly modify the dynamics. This  has already been demonstrated~\cite{Johnson:2020exp}  in computations of the spectral form factor, a useful diagnostic of the late time chaotic behaviour~\cite{Guhr:1997ve,Liu:2018hlr,Papadodimas:2015xma,Maldacena:2001kr}. (For those less familiar, an example of its classic saxophone shape over logarithmic time is given, for the (1,2) JT supergravity, in figure~\ref{fig:combined-sff-SJT}, on page~\pageref{fig:combined-sff-SJT}.) 
The late time (or large~$\beta$) piece (the ``plateau'') of that quantity  is intrinsically non--perturbative (this will be reviewed below). It can be thought of as arising from a wormhole geometry that connects the two boundaries of length $\beta$ that represent the partition function (see the righthand side of figure~\ref{fig:spectral_form_factor_diagrams} (page~\pageref{fig:spectral_form_factor_diagrams})), but with highly {\it non--perturbative} dressing. (In contrast, the geometry that controls the early part of the ``ramp'' is a cylinder diagram in perturbation theory; a perturbative wormhole.) The non--perturbative piece dominates at low temperatures, as will be shown analytically later in this paper. 

Crucially, there are   analogues of this piece when there are more boundaries, and they are naturally computed in the double--scaled matrix formalism.
This is important because for the replica method, contributions with more boundaries must be computed.  So the dominant (non--perturbative) wormholes play a leading role at low temperature, and (for small $\hbar{=}e^{-S_0}$) it is argued that an explicit result can be extracted for $\langle\ln Z(\beta) \rangle$.  Hence, the free energy $F(\beta) {=}{-}T\langle \ln Z(\beta)\rangle$ can be computed, and from it the energy, entropy, and specific heat ($U, S$ and~$C$) in this regime, with {\it e.g.,} the  result (for a toy model of (1,2) JT supergravity):
\bea
\label{eq:replica-results}
F &=& -S_0T-\sigma T^2 \ ,\quad  U = \sigma T^2\ ,\nonumber\\
 S &=& S_0+2\sigma T\ ,\quad  C = 2\sigma T\ ,
\eea
(up to an additive constant for $F$)
where $S_0$ is the extremal entropy,  and  $\sigma$ is the value of the spectral density $\rho(E)$ at $E{=}0$: $\sigma{=}\rho(0){=}\mu^2/(4\hbar^2)$, where~$\hbar{=}e^{-S_0}$.  (Here, the constant $\mu$ can be set to 1. Its meaning will be explained in the next section.) There is an analogous form suggested for JT gravity (defined as a matrix model, with a  low energy sector as discussed in footnote~\ref{foot:low-E} (page~\pageref{foot:low-E}).). There are higher order corrections in~$T$ (and~$\hbar$) that are not computed in this paper\footnote{Refs.~\cite{Okuyama:2019xbv,Okuyama:2020ncd,Okuyama:2020qpm} also use minimal models to study aspects of JT systems at  low temperature, but with  different  goals.}.  In sharp contrast to the expression~(\ref{eq:high-T-S}), the results~(\ref{eq:replica-results}) make sense all the way down to $T{=}0$.

The main lesson from these results is that there is a sensible effective thermodynamic system to be written down for these systems. It is controlled by the (single) natural low energy scale to be found in the model, which for these toy models is~$1/\rho(0)$. For the   (1,2) model, this scale is directly proportional to the location of the first bump (visible only in a non-perturbative treatment) of the density $\rho(E)$ (in the Bessel limit). From a matrix model perspective this bump can be interpreted as  the average value, $E_{\rm gap}$ of the first excited energy level of the system, which again seems a natural scale  to control the low energy dynamics. From a near-extremal black hole perspective, such a scale  naturally arises too (see {\it e.g.} refs.~\cite{Preskill:1991tb,Maldacena:1996ds,Maldacena:1998uz,Page:2000dk} and refs.~\cite{Iliesiu:2020qvm,Heydeman:2020hhw} for the more recent JT gravity context, although the physics here is at exponentially smaller temperatures than in those works). 

({\bf Note added:} Since this manuscript appeared, subsequent work~\cite{Johnson:2021rsh} using different computational methods has confirmed the quadratic fall-off of~$F(T)$, and extended it beyond just toy models of JT gravity and supergravity. That work also supports the observation that the effective energy scale controlling the curvature is correlated with the first non-perturbative bump in the density $\rho(E)$.)

 A final note is that, as already remarked, the behaviour (including $\rho(0){\neq}0$, $S(T{=}0){\neq}0$, and linear $C(T)$) shares many features with long known non--trivial disordered complex phases familiar from condensed matter physics~\cite{doi:10.1080/14786437208229210,PhysRevLett.39.41}.  Being able to model such disordered behaviour  using quantum gravity theories of JT--type could well be a new, fruitful window of opportunity to learn new physics.

\bigskip

\noindent An outline of this paper is as follows:

 {\bf Section~\ref{sec:toolbox}} is a  summary of many of the key pieces of minimal string theory that will be used in this paper. The dictionary that translates between JT--type systems and minimal string methods is still growing, and aspects of this summary (as well as results later in this paper) constitute new entries into it. The business of capturing low energy physics of a JT system is explored, including discussion about the role of the non--perturbative effects that arise in that regime. This is (in part) preparation for what is to come in later sections. 

{\bf Section~\ref{sec:JT-supergravity-results-12}} displays the results of non--perturbative studies of the (1,2) JT supergravity, including the density of states, and the spectral form factor. Some properties of $\rho(0)$, the  density of states at~$E{=}0$, are extracted. 

{\bf Section~\ref{sec:low-energy}} delves further into the low energies studies, noting along the way some interesting properties of the  system at general $E$,  elucidating and contrasting the  roles of FZZT D--branes and ZZ D--branes (as well as R--R fluxes) in   JT supergravity. The analytic treatment of the low energy (at small~$\hbar$) sector is done here (it is simply the ``Bessel'' model explored in this context in refs.~\cite{Stanford:2019vob,Johnson:2019eik}), including extracting a simple formula for~$\rho(0)$, which compares well to the full result. 

{\bf  Section~\ref{sec:low-temperature}} computes the leading free energy using the replica method, yielding the results given in equations~(\ref{eq:replica-results}), among others. Readers already familiar with the minimal model formalism and the logic of low energy limits of JT gravity can jump straight to this section. {\bf 

Section~\ref{sec:conclusions}} ends the paper with further  discussion.

\section{Double Scaled\\ Matrix Model Toolkit}

\label{sec:toolbox}
This section will summarize and discuss some of the centrepieces (from the perspective of this paper and the three leading up to it~\cite{Johnson:2019eik,Johnson:2020heh,Johnson:2020exp}) of the toolkit that  emerge from double--scaled random matrix models for use in formulating JT gravity and supergravity. It is not intended to be an exhaustive review, and nor is it entirely a review, since some of the ideas and perspective do not seem to be present in the existing literature (at least not in this context).  For older but more extensive reviews of the basic tools from a more direct string theory perspective, see {\it e.g.} refs.~\cite{Ginsparg:1993is,DiFrancesco:1993cyw,Seiberg:2004at}. (Note also that in addition to the papers mentioned above there is a growing literature of work on studying aspects of JT gravity using minimal strings, and/or  Liouville theory, starting with a suggestion in ref.~\cite{Saad:2019lba} and including {\it e.g.}, refs.~\cite{Okuyama:2019xbv,Okuyama:2020ncd,Betzios:2020nry,Mertens:2020hbs,Mertens:2020pfe,Mertens:2019tcm,Okuyama:2020qpm}.)

\subsection{The Minimal String Theories}
\label{sec:minimal-string-theories}
Certain random matrix ensembles, after the double scaling limit~\cite{Brezin:1990rb,Douglas:1990ve,Gross:1990vs,Gross:1990aw}, produce families of models that sometimes can be identified as ``minimal'' models of string theory. In some cases there  is a natural organization of the physics in terms of  a Hamiltonian~\cite{Gross:1990aw,Banks:1990df,Douglas:1990dd} for a 1D quantum mechanics: 
\be
\label{eq:schrodinger}
{\cal H}=-\hbar^2\frac{\partial^2}{\partial x^2}+u(x)\ , 
\ee
which partly follows from the fact that the model can be  solved in terms of a system of orthogonal polynomials.

The potential $u(x)$ satisfies a non--linear ordinary differential equation (ODE) called a ``string equation''.  The string equation that will occupy most of the attention of this paper is~\cite{Morris:1991cq,Dalley:1992qg,Dalley:1992br,Dalley:1992vr,Dalley:1992yi}:
 \be
\label{eq:string-equation}
u{\cal R}^2-\frac{\hbar^2}{2}{\cal R}{\cal R}^{\prime\prime}+\frac{\hbar^2}{4}({\cal R}^\prime)^2=\hbar^2\Gamma^2\ ,
\ee
where 
\be
\label{eq:flow-object}
 {\cal R} \equiv \sum_{k=1}^\infty t_k {\tilde R}_k[u] + x\ .
\ee
Here, ${\tilde R}_k[u]$ is the  $k$th order polynomial in $u(x)$ and its $x$--derivatives defined by Gel'fand and Dikii~\cite{Gelfand:1975rn}. They have a purely polynomial in $u(x)$ piece, which is $u(x)^k$,  a purely derivative linear piece, $u(x)$ $x$--differentiated $2k{-}2$ times, and then non--linear mixed terms involving $u(x)$ and its $x$--derivatives. They will not be needed in this paper, and so will not be listed. 

The string equation~(\ref{eq:string-equation}) arose from studying double--scaled complex matrix models. While the models they provided were first offered~\cite{Dalley:1992qg,Dalley:1992vr}  as alternative, better non--perturbatively defined, models of the physics captured by the bosonic $(2k{-}1,2)$ minimal string theories, they were much later~\cite{Klebanov:2003wg} identified as the $(4k,2)$ type~0A minimal string theories. Many interesting, and sometimes peculiar, properties of the equations were  uncovered in the early works, and more recently in {\it e.g.} refs.~\cite{Carlisle:2005wa,Carlisle:2005mk,Johnson:2006ux}. Several of those properties have now been shown to have a natural home in the JT gravity and supergravity context.

It  was shown in ref.~\cite{Johnson:2020heh} that if the constants $t_k$ are chosen in this particular combination:
\begin{equation}
\label{eq:teekay_SJT}
t_k=\frac{\pi^{2k}}{(k!)^2}\ ,
\end{equation}
and the solutions of the equation with the following boundary conditions for each $k$ are used:
\bea
\label{eq:boundary-conditions}
u(x)&\to&0\quad{\rm as}\quad x\to+\infty\ ,\nonumber\\
u(x)&\to&(-x)^{\frac1k}\quad {\rm as}\quad x\to-\infty\ . 
\eea
then the system defines, fully non--perturbatively, the $(2\Gamma{+}1,2)$ JT supergravity theories discussed by Stanford and Witten in ref.~\cite{Stanford:2019vob}, where the notation refers to  the $(\boldsymbol{\alpha},\boldsymbol{\beta})$   Altland--Zirnbauer~\cite{Altland:1997zz} classification\footnote{So far, particular attention was given to the cases $\Gamma{=}0,{\pm}\frac12$, but since the equation seems to naturally define smooth non--perturbative solutions for all (certainly positive) integer and half--integer $\Gamma$,  the other supergravities would seem to be naturally defined non--perturbatively by this framework. Moreover, since ref.~\cite{Carlisle:2005mk} derived an explicit  B\"acklund transformation that construct new solutions $u(x;\Gamma{\pm}1)$ from old ones $u(x;\Gamma)$, the prototype cases seed all the others quite naturally, although they are worth exploring in their own right.} of random matrix ensembles. 

Going back to the minimal string language for a moment, there is actually a duality present in these systems that can help with organizing the understanding of some JT physics to be uncovered later. These minimal string theories are formed from minimal (super)conformal field theories (with ${\hat c}{<}1$) coupled to gravity. The conformal field theory description of the gravity sector is (super)Liouville theory~\cite{Seiberg:1990eb,Maldacena:2004sn,Seiberg:2004at}. With  the Liouville field  denoted $\varphi(z)$, the target space can be taken to be (at least in part)  ${-}\infty{\leq}\varphi{\leq}{+}\infty$. The strength of the string coupling varies in this background, with the $\varphi{=}{+}\infty$ regime being strongly coupled.

\begin{figure}[h]
\centering
\includegraphics[width=0.5\textwidth]{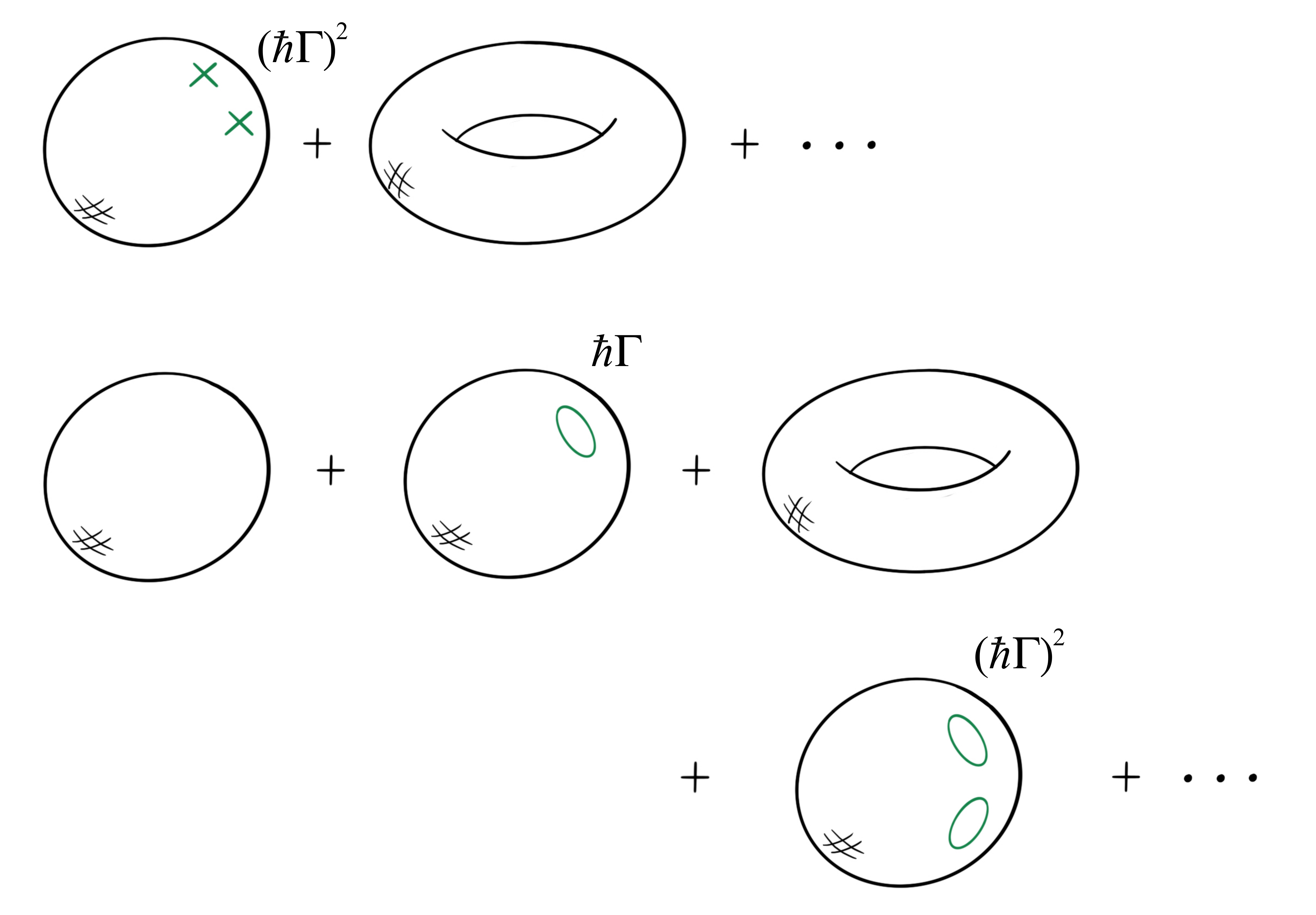}
\caption{\label{fig:world-sheets} Minimal string world--sheet diagrams corresponding to the different asymptotic regions of $x$ in equation~\ref{eq:asymptotic-expansions}. The green objects are integrated over. The loop represents $\Gamma$ D--branes, while the crosses represent pairs of closed string operator insertions building $\Gamma$ units of background R--R flux.}
\end{figure}

In these models, the spacetime physics can be readily enriched with background objects that source the R--R fields in the theory.  As string theories, there are {\it two} dual worldsheet descriptions of the spacetime physics depending upon whether $x$ is taken large and positive, or large and negative. The asymptotic expansions in these regions start out as:
\bea
\label{eq:asymptotic-expansions}
u(x)&=&0+\frac{\hbar^2\left(\Gamma^2-\frac14\right)}{x^2}+\cdots \,\,(x\to+\infty)\ ,\\
u(x)&=&(-x)^{\frac1k}+\frac{\hbar\Gamma}{k(-x)^{1-\frac{1}{2k}}}+A_k(\Gamma)\frac{\hbar^2}{x^2}+\cdots\,\,  (x\to-\infty)\ . \nonumber
\eea
where $A_k(\Gamma)$ is quadratic in $\Gamma$. (It is ${-}\Gamma^2/2$ for $k{=}1$ and $-(6\Gamma^2{+}1)/24$ for $k{=}2$.)
For $x$ large and negative, the world--sheet expansion involves closed strings and open strings, where $\Gamma$ (when integer) counts the number of background ZZ~\cite{Zamolodchikov:2001ah}  D--branes, located in the strongly coupled region at $\varphi{=}{+}\infty$. See the second two rows in figure~\ref{fig:world-sheets}.
 If $x$ is large and positive, the world--sheet expansion involves purely closed strings, with $\Gamma$ units of background R--R flux (first row in figure~\ref{fig:world-sheets}). The transition  between each asymptotic regime is an example of a geometric transition, re--summing an open--closed description into a purely closed one.  For $\Gamma$ half--integer, minimal string interpretations have not been fully worked out, but the work of ref.~\cite{Johnson:2020heh} connecting $\Gamma{=}{\pm}\frac12$ to (2,2) and (0,2) JT supergravity, where the geometries are non--orientable, suggests that the minimal strings are also non--orientable. So $\Gamma{=}n{+}\frac12$ with integer $n$ presumably involves non--orientable world--sheets and $n$ units of R--R flux, or ZZ D--branes. This would be interesting to check. 
 
There is another kind of D--brane that is naturally incorporated in these models, an FZZT~\cite{Fateev:2000ik,Teschner:2000md} D--brane, that will be discussed in section~\ref{sec:FZZT-branes}. It is intimately connected to the {\it wavefunctions} of ${\cal H}$.

Turning back to JT physics, these minimal string interpretations will of course not have the same meaning. However, they can be useful at least organizationally. The partition function of JT gravity is built from the spectrum of  ${\cal H}$ (see next subsection for a review). Since $u(x)$ is its potential,   the above knowledge about the different regions of $u(x)$ will translate into useful interpretations of the JT physics. {\it Both} regions play a role in the physics of JT supergravity (since wavefunctions spread over all $x$). But because of how the spectral density of the JT physics is constructed (see equation~(\ref{eq:spectral-density-integration})), it will be seen that the low energy sector of the theory  {\it favours the R--R flux closed string description}. 

A final remark, for this section. It was shown in ref.~\cite{Johnson:2019eik} (and explored more in ref.~\cite{Johnson:2020exp}) that the  minimal models just described can be combined in a different way (but with $\Gamma{=}0$) to give a non--perturbatively complete theory that at high energies agrees with ordinary JT gravity, but with a different, more well--behaved low energy sector. That particular construction will not feature much in this paper and so its properties will not be reviewed, although remarks will be made at various points when certain observations apply to both JT supergravity and (that definition of) ordinary JT gravity.  Additionally, some observations about the low energy sector of JT gravity (in particular, the free energy) will be made in section~\ref{sec:JT-at-low-T-and-E}.

\subsection{Constructing the partition function}
\label{sec:partition-function}

Specifically, the partition function of the JT supergravity models are built from ${\cal H}$ in this way: 
\be
\label{eq:partition-function-from-H}
\langle Z(\beta)\rangle={\rm Tr}(e^{-\beta{\cal H}}{\cal P})\ ,
\ee
where the  projection operator ${\cal P}$ is given by
\be
\label{eq:projector}
{\cal P}{\equiv}\int_{-\infty}^\mu dx\, |x\rangle \langle x| \ .
\ee
The upper limit $\mu$ will be discussed below.   In fact, equation~(\ref{eq:partition-function-from-H}) began life as an expression for the expectation value of a ``macroscopic loop''\cite{Gross:1990aw,Banks:1990df}  of fixed length ($\beta$ in this case), but ref.~\cite{Saad:2019lba} realized that JT gravity's description in terms of double--scaled matrix models implied the macroscopic loop interpretation.  

It made sense to explore~\cite{Johnson:2019eik,Johnson:2020heh} the non--perturbative implications of this connection, and extend it to JT supergravity. The core idea was that if $u(x)$ can be constructed non--perturbatively, as a solution to string equations, it supplies (through~(\ref{eq:partition-function-from-H}))  non--perturbative definitions for JT gravity and supergravity. This was demonstrated explicitly in ref.~\cite{Johnson:2020exp}.

\begin{figure}[h]
\centering
\includegraphics[width=0.5\textwidth]{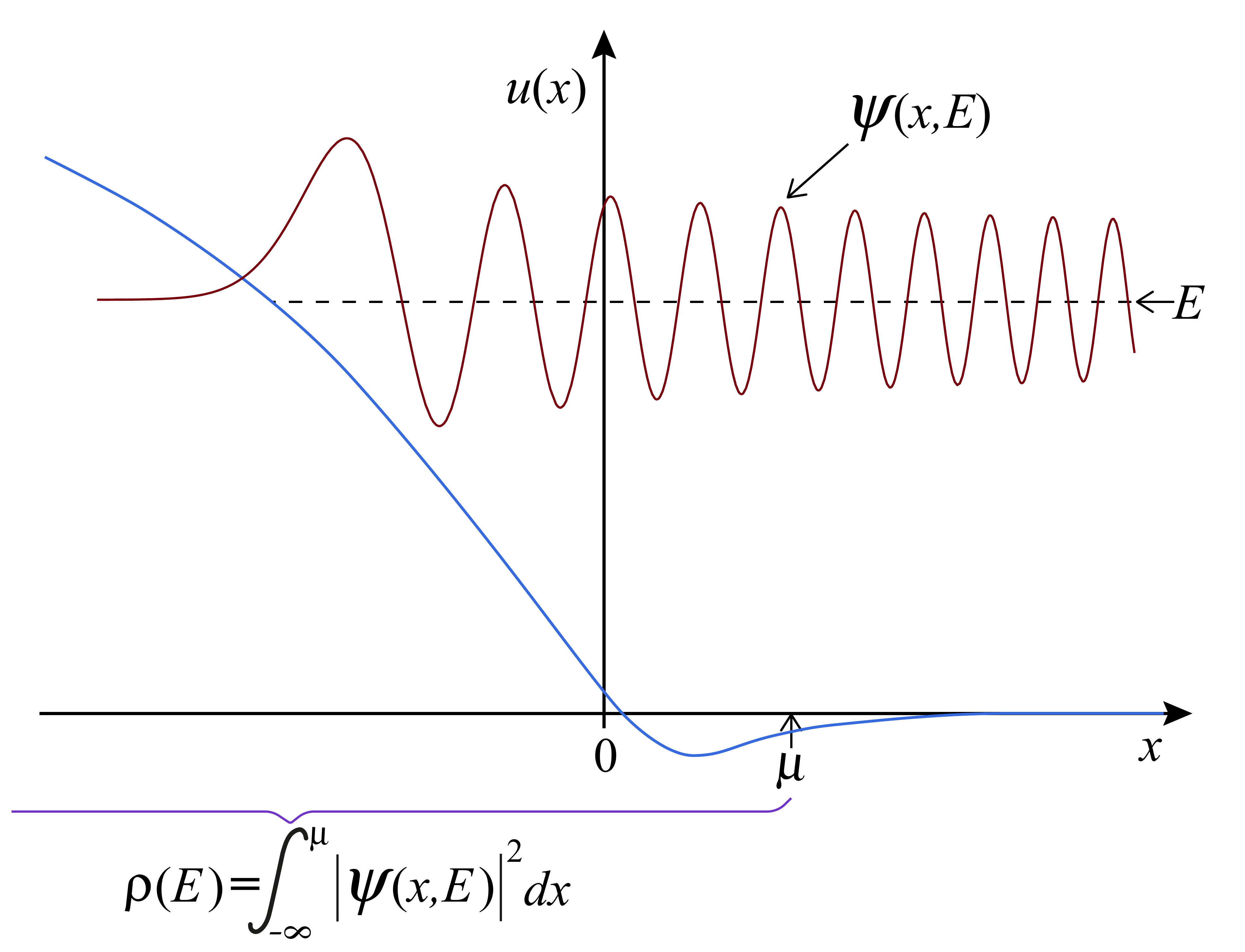}
\caption{\label{fig:potential-sketch} A visual guide to understanding how the spectral density $\rho(E)$ arises from the solution $u(x)$ of the string equation, in the case relevant to JT supergravity. A sample wavefunction is drawn at an illustrative energy $E$. The spectrum is bounded below by zero.}
\end{figure}

The partition function can be unpacked by inserting a complete set of states:
\be
\int d\psi |\psi\rangle\langle\psi | = 1\ ,
\ee
yielding
\bea
\langle Z(\beta)\rangle&=& \int_{-\infty}^\mu  dx \langle x| e^{-\beta {\cal H}} |x\rangle  \nonumber\\
             &=& \int_{-\infty}^\mu  dx \int d\psi  \langle x| e^{-\beta {\cal H}}|\psi\rangle\langle\psi | x\rangle  \nonumber\\
             &=& \int_{-\infty}^\mu  dx \int d\psi_E  \langle x |\psi_E\rangle\langle\psi_E | x\rangle e^{-\beta E} \nonumber\\
             &=&   \int dE  e^{-\beta E}  \rho(E) \ ,
             \label{eq:partition-function-integration}
\eea
where the last line reveals its description (by Laplace transform) in terms of the spectral density
\be
\label{eq:spectral-density-integration}
\rho(E) = \int_{-\infty}^\mu\!\!\! \psi(x,E)\psi^*\!(x,E) dx\ .
\ee
A visual guide to what this formula is doing may be useful here, and for future discussion. See figure~\ref{fig:potential-sketch}, where a sketch of a typical potential $u(x)$ is shown, and a sample wavefunction. 
Classically, $u{=}0$ when $x{=}0$ but it departs from this when quantum effects are turned on. The bulk of the $x$--integration over the wavefunction is from $-\infty$ to 0, but there is an extra piece coming from integrating from $0$ to $\mu$. At lower energies $E$ (adding this to the sketch is left as an exercise), the physics picks up more of the wavefunction's behaviour in the part of the potential in this latter $(0{<} x{\leq}\mu)$ neighbourhood, as compared to the exponentially small parts leaking off to the left $(x{<}0)$ region.  According to earlier comments in subsection~\ref{sec:minimal-string-theories}, the minimal string character of $u(x)$ in this region is best described in terms of closed strings with R--R fluxes. This will be discussed further in section~\ref{sec:tale-of-tails}.

Knowledge of the full spectrum of the Hamiltonian~${\cal H}$ allows a great deal about the theory to be computed. For example, ref.~\cite{Johnson:2020exp} showed explicitly how to solve the string equations (in a controlled truncation to finite order, since they are formally of infinite order) for~$u(x)$, solve for the spectrum, and then exhibit the fully non--perturbative spectral density (and hence partition function) of the $(0,2)$ and $(2,2)$ supergravity models (as well as for a proposed non--perturbative completion of ordinary JT gravity). It is straightforward to do this also for the $(1,2)$ case, but for brevity that case was not presented there. It will be presented here in  section~\ref{sec:JT-supergravity-results-12}.

The formalism outlined so far is very powerful. It allows for not just the computation of the partition function, but also correlation functions of multiple copies of it as well. This was also done (for two copies) in ref.~\cite{Johnson:2020exp}, allowing the full spectral form factor to be computed. Many useful  insights about the physics of the various supergravity models were gained from this, as well as the important issue of how this gravity computation performs what can be interpreted as an ensemble average, due to the presence of wormhole sectors.  These insights will be useful for the free energy computations to be presented later in this paper and so aspects of the two--point function computation will be reviewed and discussed in section~\ref{sec:sff-and-wormholes}. The spectral form factor for the (1,2) supergravity will be presented here in section~\ref{sec:JT-supergravity-results-12}. 

However, before doing so, and because the story will naturally be driven toward low energy, some observations about  low energy physics and  the tail of the spectral density are worth making, next.

\subsection{A Tale of Tails} 
 \label{sec:tale-of-tails}
A great deal of intuition about the physics (perturbative and non--perturbative) of double--scaled models, and hence the JT theories that can be built out of them, can be gleaned from studying the potential $u(x)$. 
In general, $u(x)$ can be separated into a classical piece $u_0(x)$ (the leading piece in the limit $\hbar{\to}0$), and  the rest,  a mixture of parts that are perturbative and non--perturbative in the small $\hbar$ expansion.  There is  the separate issue of solving for the spectrum of ${\cal H}$, for a given $u(x)$, which has its own perturbative and non--perturbative distinctions. (The two issues should not be confused.) 

Turning to low energy limits, an important hierarchical structure that arises is the notion that the low energy physics is captured by a simple model of the very end or ``tail'' of the spectrum. This happens most sharply in the limit of small $E$ {\it and} small~$\hbar$. The leading piece of the spectral density $\rho(E)$ for small~$\hbar$ is the  classical spectral density $\rhoo(E)$. In  JT gravity or JT supergravity, $\rhoo(E)$  can be expanded as a series in $E$ of the form $\sum_k A_k E^{k-\frac12}$, for $k$ integer\footnote{This discussion is for undeformed JT models. There is  recent work~\cite{Witten:2020ert,Maxfield:2020ale} on the spectral density for deformed JT gravity models.}. For JT gravity the series starts at $k{=}1$, while for JT supergravity it begins at~$k{=}0$. For low energy therefore, the dominant behaviour is given by the lowest power of $E$.  This is the tail, and its nature depends upon the system under study.

The double scaled Hermitian matrix models~\cite{Brezin:1990rb,Douglas:1990ve,Gross:1990vs,Gross:1990aw} yield the $(2k{-}1,2)$ minimal string theory models. The $k$th model  has a  spectral density with classical dependence $E^{k-\frac12}$. The $k{=}1$ model  ({\it aka} the Airy model) is  therefore the model of the  ``tail'' of the distribution, which is why it is often used as a  prototype of various low energy features of JT gravity~\cite{Saad:2019lba}.
Its potential is $u(x){=}{-}x$, and it has no corrections, perturbative or non--perturbative. The spectral problem ${\cal H}\psi{=}E\psi$ is, after a change of variables, just the defining equation for the Airy function. The resulting wavefunctions are:
\be\label{eq:airy-wavefunction}
\psi(E,x)=\hbar^{-\frac23}{\rm Ai}(-\hbar^{-\frac23}(E+x))\ ,
\ee
with a decomposition into parts that are perturbative and non--perturbative in $\hbar$.

The double--scaled complex matrix models~\cite{Morris:1991cq,Dalley:1992qg,Dalley:1992br,Dalley:1992vr,Dalley:1992yi} yield the $(4k,2)$ minimal string theory models that also have the characteristic $E^{k-\frac12}$ character (at high $E$), but each of them also has an $E^{-\frac12}$ piece (it comes from the $u(x){\to}0$ asymptotic in equation~(\ref{eq:boundary-conditions})). So while the $k{=}1$ model  might have been expected to be  the low energy tail, there is an additional sector---a special barb or tip of the expected tail, as it were---that dominates the physics\footnote{It can be thought of as a universal $k{=}0$ model, although that terminology won't be used in this paper.}. (There is a way of turning it off, in which case the $k{=}1$ model  plays  the role of the tail (see below). This happens for ref.~\cite{Johnson:2019eik}'s use of these models to give a non--perturbative definition of ordinary JT gravity.)

The physics of the tip has a model of its own, and it is the Bessel model. The potential in this case comes from the string equation~(\ref{eq:string-equation}) by setting ${\cal R}{=}{-}x$, and the resulting solution is:
\be
\label{eq:bessel-potential}
u(x){=}\frac{\hbar^2\left(\Gamma^2{-}\frac14\right)}{x^2} \ .
\ee
 After a change of variables, the  spectral problem of ${\cal H}$ with this potential can be written~\cite{Carlisle:2005mk,Johnson:2019eik} as  Bessel's equation. The wavefunctions are: 
\be
\label{eq:bessel-wavefunction}
\psi(E,x)=\frac{1}{\sqrt{2}\hbar}x^{\frac12}J_\Gamma\left(\frac{E^\frac12 x}{\hbar}\right)\ ,
\ee
again with a decomposition into perturbative and non--perturbative parts. The Bessel model is a good model of the very tip of the low energy tail for JT supergravity\footnote{This is why it played a role in some of ref.~\cite{Stanford:2019vob}'s discussion of  JT supergravity. The fact that it emerges naturally from the string equation and hence lurks at the tail of all its minimal  models was a crucial clue for the JT supergravity constructions of refs.~\cite{Johnson:2020heh}.}.
(Notice that from the point of view of minimal strings, the form of $u(x)$ in equation~(\ref{eq:bessel-potential}) is just two closed string diagrams: a torus, and a twice punctured sphere inserting~$\Gamma$ units of R--R charge, at least for $\Gamma$ integer. For $\Gamma{=}{\pm}\frac12$ the punctured sphere is presumably replaced by  a Klein bottle.)

\begin{figure}[h]
\centering
\includegraphics[width=0.48\textwidth]{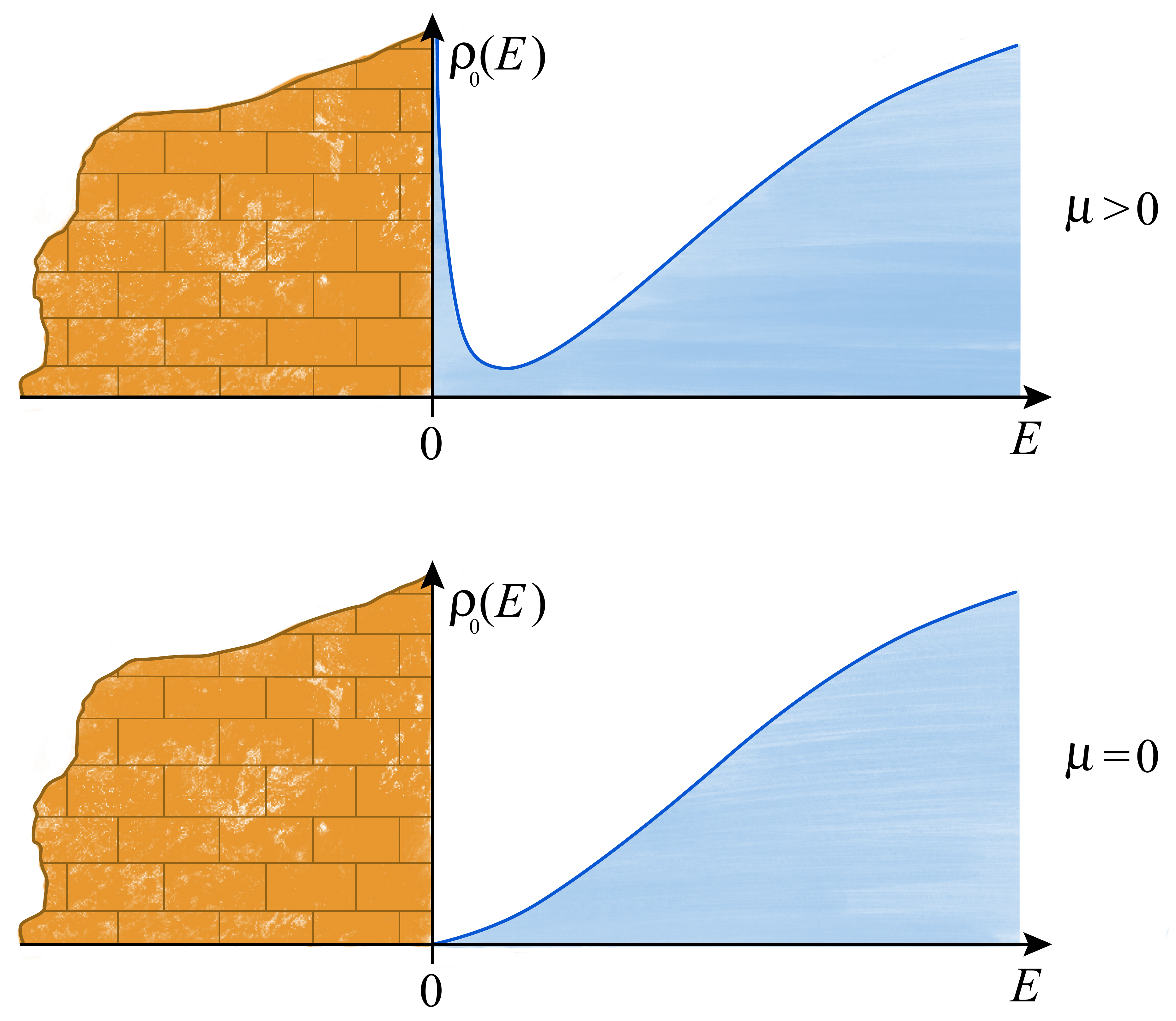}
\caption{\label{fig:the-wall} The effect of the wall on the spectral density in the classical case. If $\mu{=}0$ the divergent $1/\sqrt{E}$ meniscus goes away. When quantum effects are turned on,  a finite density at $E{=}0$ can result whether $\mu$ vanishes or not.}
\end{figure}

The presence of this special tip of the tail is controlled by the parameter $\mu$ in the projector integral~(\ref{eq:projector}), giving  the definition of the spectral density in equation~(\ref{eq:spectral-density-integration}). It is useful to think of the density integral  in two parts:
\be
\label{eq:spectral-density-integration-2}
\rho(E) = \int_{-\infty}^0 \psi(x,E)\psi^*\!(x,E) dx+\int_{0}^\mu \psi(x,E)\psi^*\!(x,E) dx\ .
\ee
At low energy and small $\hbar$, the physics comes from the second part, and $u(x)$ is given by equation~(\ref{eq:bessel-potential}) in this region, so its classical piece is zero. The contribution to the spectral density is\footnote{To see this from the expression above, expand the wave function for $\Gamma{=}0$ from equation~(\ref{eq:bessel-wavefunction}) around large $E$ (or $x$). The  leading piece is $\psi=(\pi^2\hbar^2 E)^{-\frac14}{+}\cdots$.}:  
\be
\label{eq:meniscus}
\rhoo(E)_{\rm tip} = \frac{\mu}{\pi\hbar\sqrt{E}}\ .
\ee
A picture of all this is in order. See figure~\ref{fig:the-wall}. 

The complex matrix model naturally provides a spectrum that lies on the positive $E$ line, in contrast to the Hermitian matrix model that has ${-}\infty{\leq}E{\leq}{+}\infty$. So it can be thought  of~\cite{Dalley:1992qg} as  coming with an infinite ``wall'' at $E{=}0$. 
The spectral density ends at $E{=}0$, but can be ``pushed'' into the wall by turning on non--zero (positive)~$\mu$. At the risk of overdoing the analogy, there is a sort of meniscus that results, with the density climbing up the wall. Classically, this is the divergence~(\ref{eq:meniscus}). Quantum mechanical non--perturbative corrections can conspire to reduce this divergence, or amplify it. In fact, even if a model has $\mu{=}0$ and hence  $\rho$ classically vanishes at the wall, quantum effects can generate a non--zero value  $\rho(0)$ there. This is the case for the non--perturbative definition of JT gravity presented in ref.\cite{Johnson:2019eik} (see also ref.~\cite{Johnson:2020exp}).

The standard presentation of JT supergravity~\cite{Stanford:2019vob} has $\mu{=}1$ (in the conventions of this paper), but ref.~\cite{Johnson:2020exp} showed aspects of how other values of~$\mu$ are incorporated. Changing $\mu$ is a useful probe of the structure of the theory since it naturally controls the action of instanton effects in the non--perturbative physics.  (In minimal string theory it corresponds to the coefficient of the Liouville potential.)

\subsection{Two point function, spectral form factor,\\ and the role of wormholes}
\label{sec:sff-and-wormholes}
\begin{figure}[h]
\centering
\includegraphics[width=0.5\textwidth]{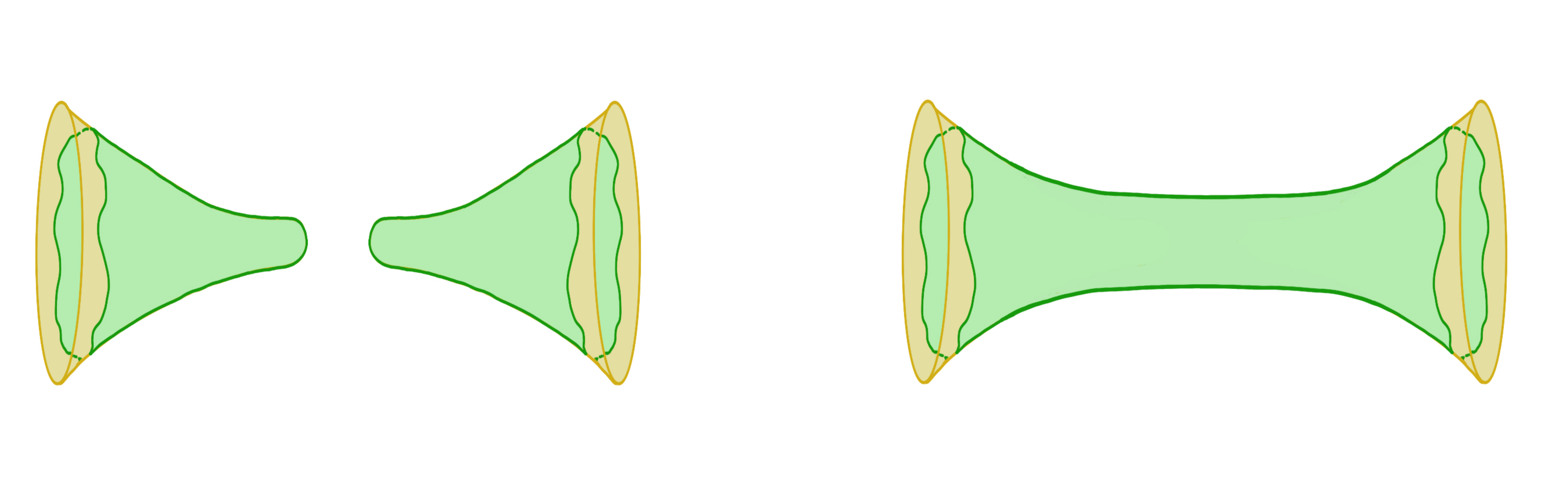}
\caption{\label{fig:spectral_form_factor_diagrams} Disconnected {\it vs.} connected (wormhole) contributions to the two--point correlator of $Z$.}
\end{figure}

As an application of the minimal string approach's access to non--perturbative JT physics,  the spectral form factor was computed for the full JT gravity and supergravity models in ref.~\cite{Johnson:2020exp}. This follows from computing the correlator of two copies of the partition function. There is a disconnected piece, $\langle Z(\beta)\rangle \langle Z(\beta^\prime)\rangle $ and a connected piece. See figure~\ref{fig:spectral_form_factor_diagrams}.   

\noindent The  connected part  is given by~\cite{Banks:1990df}:
\bea
\label{eq:two-point-connected}
\langle Z(\beta) Z(\beta^\prime)\rangle &=& {\rm Tr}(e^{-\beta{\cal H}}(1-{\cal P}) e^{-\beta^\prime{\cal H}}{\cal P})\ .
\eea
It can be unpacked with similar manipulations to those done in equation~(\ref{eq:partition-function-integration}):
\begin{eqnarray}
\label{eq:correlator-connected}
\langle Z(\beta) Z(\beta^\prime)\rangle 
&=& {\rm Tr}(e^{-(\beta+\beta^\prime){\cal H}}{\cal P}) - {\rm Tr}(e^{-\beta{\cal H}}{\cal P} e^{-\beta^\prime{\cal H}}{\cal P})\nonumber \\ 
&&\hskip-2.2cm = \langle Z(\beta{+}\beta^\prime) \rangle-\!\! \int \!\!\!dE\, e^{-\beta E}\!\int \!\!\!dE^\prime e^{-\beta^\prime E^\prime}\rho(E,E^\prime) \rho^*\!(E^\prime,E)
\ ,\nonumber
\end{eqnarray}
where 
\be
\rho(E,E^\prime) = \int_{-\infty}^\mu\!\!  dx \, \psi^*\!(x,E)\psi(x,E^\prime) \ .
\ee
It  is worth a pause to develop a little intuition for this, since it will be useful later. Setting $\beta^\prime{=}\beta$,  consider the relative strengths of the connected and disconnected pieces as a function of $\beta$.  A single copy of the partition function  has an inverse $\beta$ dependence $Z{\sim}\beta^{-p}$ (classically, $p{=}\frac12$ for JT supergravity and  $\frac32$ for ordinary JT). Squaring this to get the disconnected two point function gives~$\beta^{-2p}$. The leading positive piece of the connected part is actually $\langle Z(2\beta)\rangle$, and so its dependence is still~$\beta^{-p}$. This means that at small $\beta$ the disconnected piece will dominate the physics, reducing in size as~$\beta$ grows. It eventually falls below the size of the leading connected piece, which falls off more slowly with $\beta$. At large $\beta$ the leading connected piece  will therefore dominate. Generically there is a transition between the (total) connected and disconnected pieces  at some $\beta_{\rm c}$, the value of which depends upon $\hbar$ (if perturbative corrections are included) and the details of the theory in question. Similar considerations apply when  non--perturbative contributions are taken into account, as shown in ref.~\cite{Johnson:2020exp}.

The spectral form factor is the sum of the connected and disconnected pieces, with the substitution $\beta{\to}\beta{+}it$ and $\beta^\prime{\to}\beta{-}it$.  Some of the key features of its time ($t$) dependence (see {\it e.g.} figure~\ref{fig:combined-sff-SJT}) is controlled by the   large $\beta$ observations made above.  There is an early period where it slopes downward as the disconnected piece falls, followed by the dip where it begins to rise again due to the growing dominance of the connected part, up a ramp part (which isn't always linear---see ref.~\cite{Johnson:2020exp}), finally settling to the saturation value $\langle Z(2\beta)\rangle$ at late times, the plateau. The relative durations of these  different epochs depends upon the starting value of $\beta$ relative to $\beta_{\rm c}$, the value of~$\hbar$, and the details of the non--perturbative physics. This organization of the late time (or large $\beta$) physics can be thought of as the domination of the large $\beta$ (low temperature/energy) physics by a wormhole that connects the two nearly AdS$_2$ geometries. This, combined with the fact that the quantum gravity path integral sums over all surfaces that connect those boundary topologies, can be interpreted as not just the spectral form factor for a representative of the ``dual'' boundary theory, but the average over the whole ensemble~\cite{Cotler:2016fpe,Saad:2018bqo,Saad:2019lba}.

The presence of the wormhole in the calculus should not necessarily be thought of, in and of itself,  as a non--perturbative effect.  It is important to distinguish between its presence in an amplitude and the effects of actual non--perturbative physics. The wormholes are more properly thought of as a channel through which the non--perturbative physics, which is increasingly important at  low energy, can be communicated to important  quantities of interest ({\it  e.g.,} the spectral form factor, or the free energy, as will emerge later).

For example, {\it crucially}, the nature of the non--perturbative physics at low energy is important for the details of the features of the plateau and the transition to it from the ramp. This was elucidated in some detail in ref.~\cite{Johnson:2020exp}. This is because saturation to the ramp depends upon the dying out of late time correlations between low--lying states. The more low--lying states there are, the longer it takes for the connected piece to saturate to its ramp value. This makes, for example, the $(0,2)$ supergravity theory quite different from the $(2,2)$ theory, since while they have the same spectral density classically, with a $1/\sqrt{E}$ low energy divergence, the $(0,2)$ theory retains such a divergence after non--perturbative effects are included, while the $(2,2)$ theory's non--perturbative completion exactly cancels the spectral density to zero at $E{=}0$. As a result, saturation of the form factor is much more swift in the latter case, while it is asymptotic for the former. The same wormhole structure is present for each theory, but the distinct  non--perturbative natures of their spectra gave a different overall result. 

The $(1,2)$ JT supergravity then becomes an interesting intermediate case, since (see  section~\ref{sec:JT-supergravity-results-12}), the non--perturbative effects leave it with a finite but non--zero spectral density at $E{=}0$, so its characteristic saturation time to the plateau value will be larger than the $(2,2)$ case, but less than that of the $(0,2)$ case. This is borne out by the computations. Since this is really a low energy issue, a lot of insight can be gained by  focussing directly on the neighbourhood of  $E{=}0$, and this will be done in later sections. 

A key lesson from all this for later use in this paper is the following. When $\beta{=}\beta^\prime$, the division of the connected correlator~(\ref{eq:correlator-connected}) into the two parts has a curious form. The first term is essentially a single partition function at $2\beta$, which is why it wins at large $\beta$ over all other contributions, connected or disconnected. At large $\beta$ it can be thought of as an object in its own right, a sort of non--perturbatively dressed wormhole that becomes the most important contribution. This will persist in diagrams with more insertions of $Z(\beta)$ and will be a key simplifying feature in the replica computation of the free energy in section~\ref{sec:low-temperature}.

\section{The (1,2) JT supergravity}

\label{sec:JT-supergravity-results-12}
\subsection{Non--perturbative Potential}

The methods of ref.~\cite{Johnson:2020exp} were developed to show that explicit non--perturbative results can be extracted from the definition of JT supergravity and gravity in terms of minimal string theories, despite the fact that the defining string equation is highly non--linear and formally of infinite order. The key point was that there is a well--defined sense in which a controlled truncation of the equation can be made such that the results for the spectrum are accurate up to a given desired energy.  A truncation resulting in the solving of  13th order ODEs was done for studying JT supergravity cases (0,2) and (2,2). Similar work can be done for the (1,2) model, and the results are presented here without much additional discussion of the methods, since they are as in ref.~\cite{Johnson:2020exp}.

\begin{figure}[h]
\centering
\includegraphics[width=0.45\textwidth]{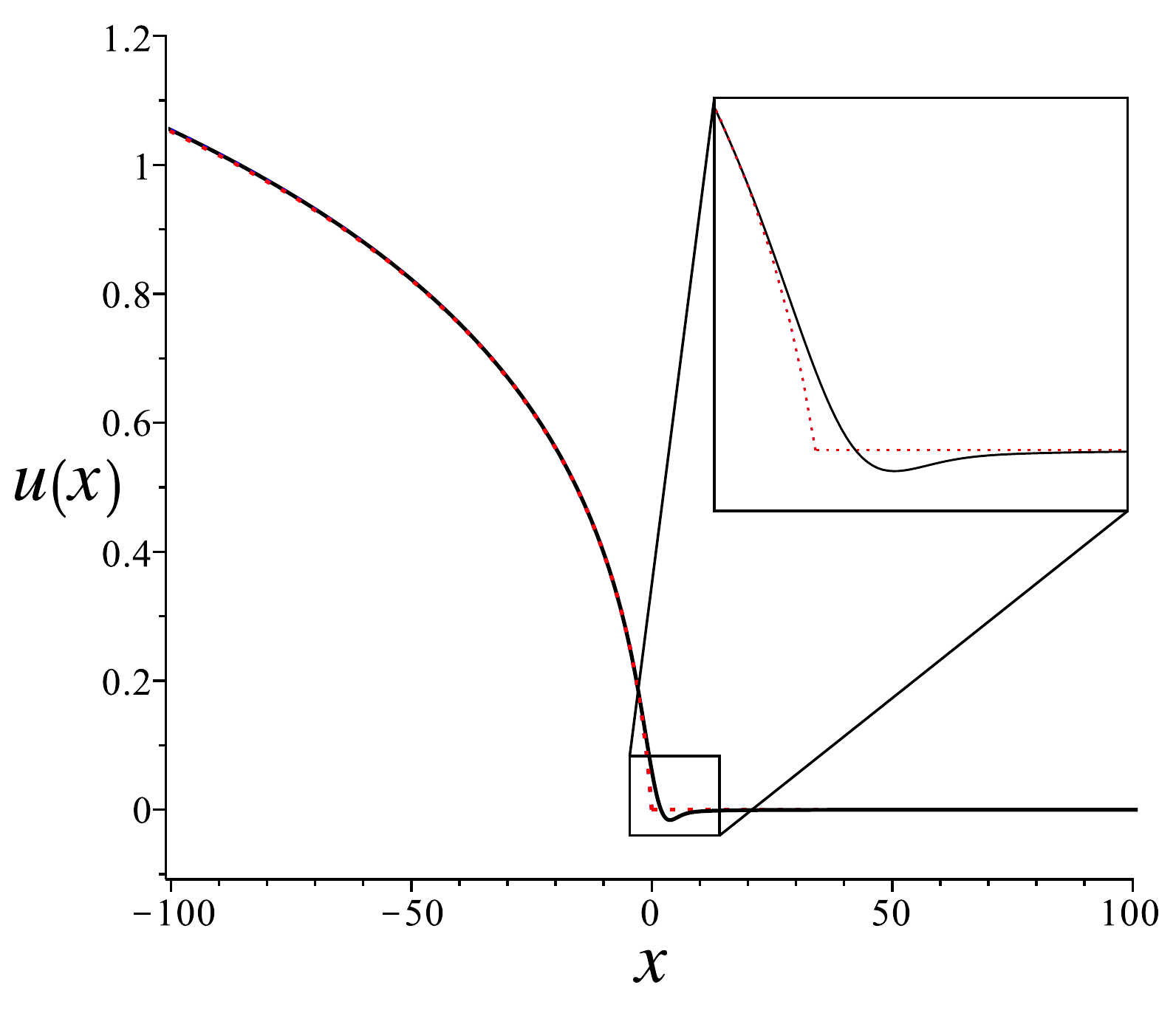}
\caption{\label{fig:truncation-examples-E} The  solution (solid line) of the string equation for truncation up to $t_6$. The inset shows the well that developed in the interior.  The full classical solution  is shown too (dotted), showing good agreement up to energy $E{\simeq}1.0$ (and somewhat beyond).}
\end{figure}

The difference is that the equations must be solved with $\Gamma{=}0$ instead of $\Gamma{=}{\pm}\frac12$. This turns out to be a harder situation to solve for two reasons. The first is that the $x{\to}{+}\infty$ boundary condition is less trivial now, and so non--zero terms must be specified in the boundary value problem.  (For $\Gamma{=}{\pm}\frac12$ the magical result that every term in the asymptotic expansion of $u(x)$ vanishes in that regime led to a certain simplicity.) The second reason is that the resulting solution has a more appreciably sized well in the interior, so instabilities can easily result in the numerical boundary value routine failing to navigate this feature.  The  trick to avoiding that was similar to what was done when solving for $\Gamma{=}{-}\frac12$ potential. Solving for $\Gamma{=}{+}\frac12$ was easier, and then the B\"acklund transformation of ref.~\cite{Carlisle:2005mk} that gives solutions differing by unit $\Gamma$ was used to  generate the $\Gamma{=}{-}\frac12$ case. Similarly, here the $u(x)$ solution for $\Gamma{=}1$ was found, and then the desired $\Gamma{=}0$ solution was built from it. It is shown in figure~\ref{fig:truncation-examples-E}.  

\subsection{Non--perturbative Spectral Density}
\label{sec:full-spectral-density}
\begin{figure}[h]
\centering
\includegraphics[width=0.45\textwidth]{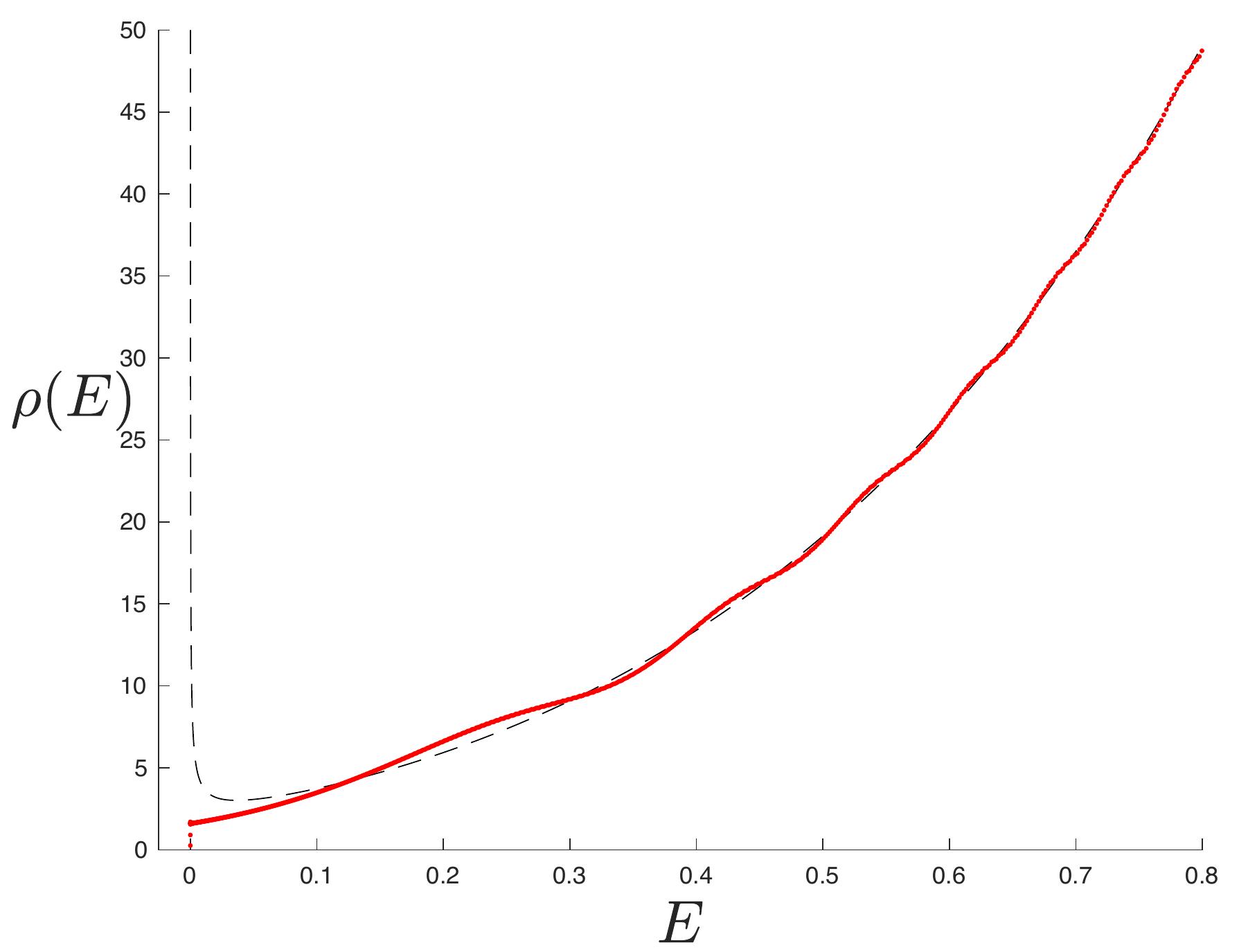}
\caption{\label{fig:non-pert-sd4} The full spectral density for the $(1,2)$ JT supergravity. The dashed  line is the disc level Schwarzian result~(\ref{eq:SJT-schwarzian}).}
\end{figure}

The next step is  solving the spectral problem of ${\cal H}$ with this potential. The same techniques described in ref.~\cite{Johnson:2020exp} were used, with no modifications. The result of doing the integral~(\ref{eq:spectral-density-integration}) to get $\rho(E)$ is presented in figure~\ref{fig:non-pert-sd4}. 
There are similarities with the full spectral densities found for the (0,2) and (2,2) cases, but there is a key difference. There is a finite and non--zero spectral density at $E{=}0$. Recall that the classical result (in this paper's normalization),
\be
\label{eq:SJT-schwarzian}
\rhoo(E)=\frac{\cosh(2\pi\sqrt{E})}{\hbar\pi\sqrt{E}} \quad{\rm (SJT)}
\ee
 is  infinite  there. The full quantum corrections cancelled it to zero for the (2,2) case, and in this case, the quantum effects have produced a  {\it finite} density, $\rho(0)$  which is (from some points of view) more  remarkable.  It is also the case that ordinary JT gravity, when given the non--perturbative definition of ref.~\cite{Johnson:2019eik}, also has  $\rho(0){\neq}0$ (it is studied more in ref.~\cite{Johnson:2020exp}). It is an example of a $\mu{=}0$ case that generates  $\rho(0){\neq}0$ from purely quantum effects.

 A goal of this paper will be to understand more about the physics of this finite density appearing at zero energy. Determining the dependence of $\rho(0)$ on the parameter~$\mu$ will be helpful in this endeavour (when comparing to some analytic results to be derived later in a special limit). This is straightforward to extract by repeating what was done to obtain figure~\ref{fig:non-pert-sd4}, but  at different $\mu$ values, reading off the value at $E{=}0$ (being careful to avoid some easy to navigate  numerical jitter in this region). 

\begin{figure}[h]
\centering
\includegraphics[width=0.45\textwidth]{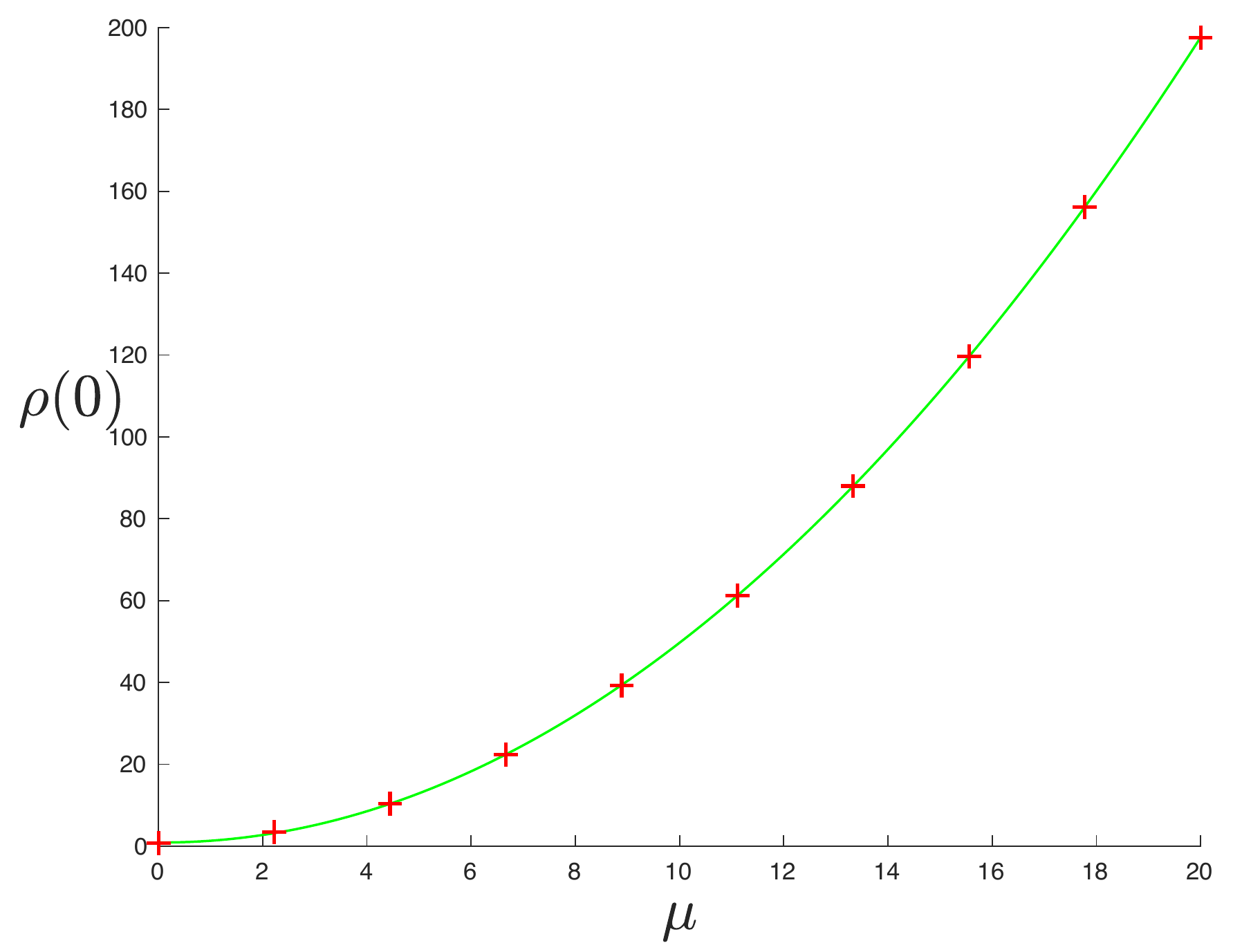}
\caption{\label{fig:mu-dependence} The crosses show the $\mu$ dependence of  $\rho(0)$  for the $(1,2)$ JT supergravity for $\hbar{=}1$. The solid line is the curve that behaves as $\rho(0){\simeq}0.4956\mu^2$ for large $\mu$. Note that $\rho(0){\neq}0$ at $\mu{=}0$. (See text.).} 
\end{figure}
The result is a clear quadratic dependence as~$\mu$ grows large, as shown in figure~\ref{fig:mu-dependence}, close to $\mu^2/2$.  
 (The normalization of ref.~\cite{Johnson:2020exp}, and hence of the results of this section, are such that there is a factor of two relationship between the values of $\mu^2$. Therefore this dependence is to be read as $\mu^2/4$, for later comparison.) Note that the dependence is not purely quadratic,  only becoming so at large $\mu$: at $\mu{=}0$ there is still a small non--zero piece. The meaning of all this will be explored further in subsection~\ref{sec:low-energy-tail-tip}, by  constructing the $E{=}0$ wavefunction.

\subsection{Spectral Form Factor}
\label{sec:spectral-form-factor}
\begin{figure}[h]
\centering
\includegraphics[width=0.48\textwidth]{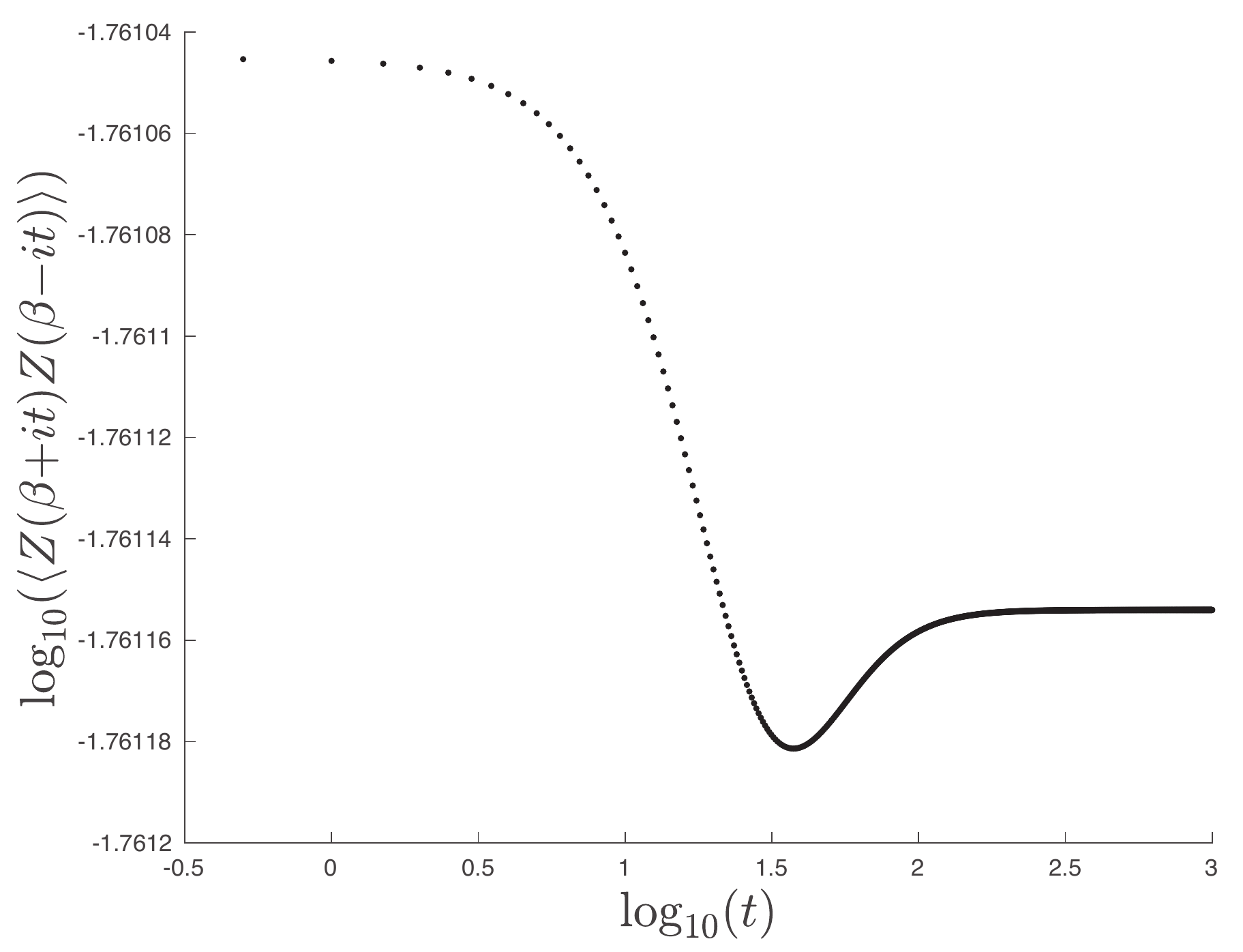}
\caption{\label{fig:combined-sff-SJT} The full  (1,2) JT supergravity  spectral form factor  {\it vs.}~$t$, at $\beta{=}50$ and $\hbar{=}1$. The key features are the ``slope" (the initial plunge), ``dip", ``ramp", and ``plateau".}
\end{figure}

The spectral form factor is the next natural step. Solving the spectrum of ${\cal H}$ resulted in a library of $\sim$875 wavefunctions and energies. Those can be used in equation~(\ref{eq:correlator-connected}) to compute the connected piece, and the last subsection already computed the partition function for use in the disconnected piece. The continuation to include $t$ is easily done numerically as before~\cite{Johnson:2020exp}, and for completeness, sample  results for the 
disconnected and connected parts are displayed in Appendix~\ref{sec:extra-stuff} in figures~\ref{fig:disconnected-sff-SJT} and~\ref{fig:connected-sff-SJT} respectively. The 
sum of these parts, constituting the whole spectral form factor, is given in below figure~\ref{fig:combined-sff-SJT}. It is interesting to compare these three figures to those obtained for the (2,2) and (0,2) supergravities. As expected, the differences are consistent with the key low energy (non--perturbative) features observed in the respective spectral densities.

\section{Low  Energy JT Supergravity}
\label{sec:low-energy}
The last section uncovered an interesting  and novel quantum feature for the (1,2) supergravity model, a finite density at $E{=}0$. It would be interesting to understand the physics of this for several reasons. First and foremost, while the other two supergravities have time--reversal invariance, this supergravity theory does not, and so is more akin to ordinary JT gravity that the others. Second, it is an intermediate situation between having an infinite $E{=}0$ density ($(0,2)$ case) and zero density ($(2,2)$ case). So it will be useful to contrast the three cases all in the same setting. This section will do that.

First, it is useful to start with some observations that are true for any $E$. Begin by writing $\cal H$ as a product:
\bea
\label{eq:product-hamiltonian}
{\cal H} &\equiv& -\hbar^2\frac{\partial^2}{\partial x^2} +u(x) \nonumber\\
&=&{\cal A}^\dagger{\cal A}
= \left(-\hbar\frac{\partial}{\partial x}+v(x)\right)\cdot \left(\hbar\frac{\partial}{\partial x}+v(x)\right)\ , 
\eea
where $u{=}v^2{-}v^\prime$. (As usual a prime denotes an $x$--derivative times $\hbar$ here.) 
Now, ref.~\cite{Dalley:1992br} showed that if $u(x)$ satisfies the string equation~(\ref{eq:string-equation}) with constant $\Gamma$, then $v$ satisfies  the following equation:
\be
\label{eq:painleveII-H}
\sum_{k=1}^\infty t_k\left\{\frac12 {\tilde R}^\prime_k[v^2-v^\prime]-v{\tilde R}_k[v^2-v]\right\}+xv+\hbar C=0\ ,
\ee
and the ${\tilde R}_k[u]$ are the Gel'fand--Dikii differential polynomials mentioned earlier (see below~(\ref{eq:flow-object})). The constant~$C$ is given by $C{=}\frac12{\pm}\Gamma$.
This is actually the Painlev\`e~II heirarchy of equations derived for double--scaled unitary matrix models~\cite{Periwal:1990gf,Periwal:1990qb,Crnkovic:1990ms,Gross:1991aj,Minahan:1991pv}. This one--to--one map of the solutions of double--scaled unitary matrix models and those of double--scaled complex matrix models  demonstrated that unitary matrix models described theories of 2D gravity\footnote{In fact, the definition of refs.~\cite{Johnson:2020heh,Johnson:2020exp} of JT supergravity in terms of complex matrix models could well be re--written in terms of these equivalent minimal models, now these days more commonly referred to as the Brezin--Gross--Witten hierarchy. A recent paper~\cite{Okuyama:2020qpm} has actually done this very re--writing.}. However, this is not the purpose of recalling these structures.

Imagine starting with a particular solution~$v(x)$, with either choice of  $C$. Now form instead  the combination ${\bar u}{=}v^2{+}v^\prime$ (note the plus sign). This is a different function than the original,~$u(x)$, and in fact it results from the alternative  factorization with ${\cal A}$ and ${\cal A}^\dagger$ reversed: $\overline {\cal H}{=}{\cal A} {\cal A}^\dagger$. In fact, this new potential ${\bar u}(x)$ satisfies the original string equation but for $\Gamma$ shifted by either $+1$ or $-1$, depending upon whether the $v(x)$ used had been chosen with  $C{=}\frac12{+}\Gamma$ or $C{=}\frac12{-}\Gamma$. This is easy to see based on the fact that the hierarchy is unchanged under the symmetry $v{\to}{-}v$ combined with $C{\to}{-}C$.

The point is that there is an important symmetry between spectra of the Hamiltonians, away from the $E{=}0$ sector. The eigenvalue $E$ equation of ${\cal A}^\dagger{\cal A}$, with eigenfunction $\psi$ supplies (by acting on the left with ${\cal A}$), an eigenfunction ${\cal A}\psi$ of ${\cal A}{\cal A}^\dagger$ with the same energy. The $E{=}0$ states are not mapped to each other though.  This system actually has the formal structure of a supersymmetric quantum mechanics (for a useful review, see ref.~\cite{Cooper:1994eh}), and any two Hamiltonians for values of $\Gamma$ separated by unity are {\it potentially} (see below) superpartner pairs.  


\subsection{Generic $E{=}0$}
\label{sec:full-E-zero}

It is important to consider the  $E{=}0$ sector, with the original ${\cal H}$ in equation~(\ref{eq:product-hamiltonian}). Something rather special happens there that was noticed a while ago in ref.\cite{Carlisle:2005wa}, and this JT context may well supply it with a natural application.  Since  ${\cal H}\psi{=}{\cal A}^\dagger{\cal A}\psi{=}0$, the wavefunction can be written, up to a normalization to be fixed later, as: 
\be
\label{eq:ground-state-wavefunction}
\psi(x) \sim \exp\left\{-\hbar^{-1}\int^x v(x') dx'\right\}\ . 
\ee
So the zero energy non--perturbative wavefunctions of the JT supergravity can be {\it fully constructed} from solutions of equation~(\ref{eq:painleveII-H}), with the $t_k$ given as in equation~(\ref{eq:teekay_SJT}).  There is a  truncation method (analogous to that used in ref.~\cite{Johnson:2020exp}) to be pursued here, with  numerical methods. Using the resulting $v(x)$ in equation~(\ref{eq:ground-state-wavefunction}) will yield  a wavefunction that when used in equation~(\ref{eq:spectral-density-integration}) to form the spectral density at $E{=}0$ will presumably yield the constant seen in figure~\ref{fig:non-pert-sd4}. This would be an interesting numerical exploration to carry out\footnote{A warm--up version was done in ref.~\cite{Johnson:2019eik}, just for the $k{=}1$ model.}.

More generally, it is worth further investigating this role that the Painlev\'e~II hierarchy has in organising the structure of the $E{=}0$ sector of JT supergravity. 

\subsection{Wavefunctions and FZZT D--branes}
\label{sec:FZZT-branes}
Before moving on, note that there is a useful  interpretation of the wavefunctions from minimal string language that should be helpful for organising thoughts in this JT context. The wavefunctions (for any $E$) are actually partition functions for probe FZZT~\cite{Fateev:2000ik,Teschner:2000md}  D--branes. Recall from section~\ref{sec:minimal-string-theories} that the Liouville field $\varphi$ part of the target space runs from $\varphi{=}{-}\infty$ a region of weak coupling, to $\varphi{=}{+}\infty$, a strongly coupled region.  The $\Gamma$ target space ZZ D--branes (or the R--R flux sources) are localized in this region. Instead\cite{Maldacena:2004sn}, the FZZT  D--branes run along the~$\varphi$ direction from ${-}\infty$, ending at a position $\varphi_0{\simeq}{-}\ln{E}$. It was shown in ref.~\cite{Carlisle:2005wa} that it is still useful to write the wavefunction  in the form~(\ref{eq:ground-state-wavefunction}), but now with $v(x)$ related to $u(x)$ by $u{-}E{=}v^2{-}v^\prime$. The resulting asymptotic expansions for $v(x){=}{-}\hbar \psi^\prime/\psi$ can be interpreted as the various diagrams that can be drawn with a boundary representing the FZZT brane, and either boundaries representing the background ZZ--branes or the R--R flux insertions. Again, the two different asymptotic regions for $x$ give one or the other.

As mentioned in section~\ref{sec:partition-function}, since the $x$--integration to construct the JT spectral density runs from ${-}\infty$ some some finite positive value $\mu$, and since wavefunctions at some given energy $E$ eventually exponentially decay to the left, at very low energies most of the contribution of the wavefunctions will come the part of the wavefunction that extends in the $x{>}0$ region. This is the region that, in minimal string terms, is best understood in terms of background R--R fluxes. This will be helpful in interpreting the low energy results to come later.

\subsection{$E{=}0$: Toward small $\hbar$}
\label{sec:Painleve-II}
What has been done in this section so far is the general $E{=}0$ case, for arbitrary $\hbar$.  As mentioned in section~\ref{sec:tale-of-tails} a simpler low energy tail can emerge by working at small~$\hbar$. An intermediate step along the way is to take the $k{=}1$ model. In that case, the most relevant  equation emerging from the system~(\ref{eq:painleveII-H}) is the lowest member of the hierarchy, the famous Painlev\'e~II equation itself:
\be
\label{eq:painleveII}
\frac12 v^{''}-v^3+xv+\hbar C = 0\ .
\ee
A unique solution to this equation with the boundary conditions needed is known to exist~\cite{Hastings1980}. It is easy to see from the equation that the leading behaviour for $v(x)$ in the negative $x$ regime is $v(x){=}{-}\hbar C/x{+}\cdots$, and so the ground state wavefunction there is, asymptotically:
\be
\label{eq:asymptotic-wavefunction}
\psi(x)\sim x^{\frac12\pm\Gamma}+\cdots\ .
\ee
On the other hand, the leading behaviour in the ${+}x$ direction for the $k$th case is $v{\sim}x^{1/2}{+}\hbar C/2x+\cdots$, 
which gives:
\be
\psi(x)\sim x^{\frac12(\frac12\pm\Gamma)}\exp\left(-\frac23\frac{x^\frac32}{\hbar}\right)
+\cdots .
\ee This shows that the ground state wavefunctions are non--normalizable. This means that the putative supersymmetry observed earlier is  actually broken, which makes sense in this context. Nevertheless, the mapping structure remains as a solution generating symmetry\cite{Carlisle:2005mk}, allowing  $u(x)$ for a given value of $\Gamma$ to be turned into $u(x)$ for $\Gamma\pm1$. It could be interesting to explore whether this (broken) supersymmetric structure, combined with the underlying integrable hierarchies,  can be used to learn anything more about JT supergravity.

\subsection{$E{=}0$: The Tip of the Tail}
\label{sec:low-energy-tail-tip}
As mentioned in section~\ref{sec:tale-of-tails}, the $k{=}1$ tail just studied has an extra piece that dominates the low energy physics in the small $\hbar$ limit. The full model of it is the Bessel model with the exact wavefunctions given in equation~(\ref{eq:bessel-wavefunction}), where the normalization in now in place. In the projection integral and spectral density integration, the range of $x$ has reduced to $0{\leq}x{\leq}\mu)$ (so only the second term in equation~(\ref{eq:spectral-density-integration-2}) remains). For the $E{=}0$ case, the case $\Gamma{=}{-}\frac12$ is divergent, giving a divergent $E{=}0$ spectral density. The case $\Gamma{=}{+}\frac12$ vanishes at $E{=}0$, with corresponding vanishing density. 

The case $\Gamma{=}0$ is special since $J_1(0){=}1$. In this case, the resulting wavefunction $\psi{=}(\sqrt{2}h)^{-1}x^\frac12$ (note the agreement with equation~(\ref{eq:asymptotic-wavefunction})) gives: 
\be
\label{eq:special-density}
\rho(0)=\frac{\mu^2}{4\hbar^2}\ .
\ee 
It is worth emphasizing again that this is the result for the case of small $\hbar$. It will not numerically match the result obtained in section~\ref{sec:full-spectral-density} for the full (1,2) JT supergravity theory at $\hbar{=}1$. In that regime, the appropriate full wavefunctions are those described in equation~(\ref{eq:ground-state-wavefunction}). Here, instead, is the result in a limit where much can be studied analytically, but the lessons learned will apply to the $E{=}0$ sector for more general $\hbar$. 
For example, the~$\mu^2$ dependence is interesting. It should not be exactly quadratic at arbitrary $\hbar$, since the $x$ dependence of the wavefunctions is more complicated than just $x^\frac12$, but the leading dependence will be of this form, and so it should be expected that a quadratic dependence  will indeed settle in at large enough  $\mu$. Indeed, this was confirmed  with the explorations done in the full model in subsection~\ref{sec:full-spectral-density}. 

To further study  the consequences of this $E{=}0$ density on other physics, an  option is to examine the thermal properties of the system in the neighbourhood of $E{=}0$. This will be done in the next  section. In preparation, it is prudent to expand the wavefunctions in low energy (and effectively small $x$), keeping only the first few leading terms. The expansion to use in the wavefunction expression~(\ref{eq:bessel-wavefunction}) for the Bessel functions is, for small $z$:
\be
\label{eq:expand-J}
J_\Gamma(z) =\sum_{m=0}^\infty\frac{(-1)^m}{m!{\widetilde{\Gamma}}(\Gamma+m+1)}\left(\frac{z}{2}\right)^{\Gamma+2m}\ , 
\ee
where  ${\widetilde\Gamma}$ means the $\Gamma$--function, an unfortunate necessity since the symbol $\Gamma$ is already in use. The result is:
\bea
\label{eq:wave-function-small-E}
\psi&\simeq&\frac{1}{\sqrt{2}\hbar}\cdot  \frac{x^{\frac12+\Gamma}}{{\widetilde\Gamma}(\Gamma+1)}\left(\frac{E^\frac12}{2\hbar}\right)^\Gamma\biggl(1-\frac{1}{4(\Gamma+1)}\frac{x^2 E}{\hbar^2}\nonumber\\
&&\hskip1.5cm+\frac{1}{32(\Gamma+1)(\Gamma+2)}\frac{x^4 E^2}{\hbar^4} + \cdots\biggr)\ .
\eea
For later discussion, it is useful to write explicitly some of the $(1,2)$ ({\it i.e.,} $\Gamma{=}0$) case:
\be
\label{eq:wave-function-small-E-12}
\psi\simeq \frac{1}{\sqrt{2}\hbar}\cdot x^\frac12\left(1-\frac14\frac{x^2 E}{\hbar^2}+\frac{1}{64}\frac{x^4 E^2}{\hbar^4}+ \cdots\right) \ ,
\ee
while for $(2,2)$ ({\it i.e.,} $\Gamma{=}{+}\frac12$):
\be
\label{eq:wave-function-small-E-22}
\psi\simeq \frac{1}{\sqrt{2} \hbar} \cdot \sqrt{\frac{2}{\pi}}\frac{E^\frac14}{\hbar^\frac12}x\left(1-\frac16\frac{ x^2 E}{\hbar^2} + \frac{1}{120}\frac{x^4 E^2}{\hbar^4}+\cdots\right) \ ,
\ee
and for $(0,2)$ ({\it i.e.,} $\Gamma{=}{-}\frac12$):
\be
\label{eq:wave-function-small-E-02}
\psi\simeq \frac{1}{\sqrt{2} \hbar} \cdot  \sqrt{\frac{2}{\pi}}\frac{\hbar^\frac12}{E^\frac14}\left(1-\frac12\frac{ x^2 E}{\hbar^2}+ \frac{1}{24}\frac{x^4 E^2}{\hbar^4}+\cdots\right) \ .
\ee
Notice that these expansions are in the opposite regime to what is usually used to get classical (large $E$) physics. Also, from the minimal string perspective they can  be interpreted as partition functions of FZZT probe branes in  closed string backgrounds (with $\Gamma$ units of R--R flux, for integer $\Gamma$), as mentioned in subsection~\ref{sec:FZZT-branes}. It is instructive to construct $v(x){=}{-}\hbar\psi^\prime/\psi$  for these Bessel cases, yielding: 
\bea
\label{eq:probe-fzzt-general}
v(x) &\simeq& -\frac{\hbar}{x}\biggl(\Gamma+\frac12-\frac12\frac{1}{(\Gamma+1)}\frac{x^2E}{\hbar^2}\nonumber\\
&&\hskip1.0cm-\frac{1}{8}\frac{1}{(\Gamma+1)^2(\Gamma+2)}\frac{x^4E^2}{\hbar^4} +\cdots\biggr)\ ,
\eea
%
%
The leading term is the $-\hbar C/x$ seen earlier (see {\it e.g.,} above equation~(\ref{eq:asymptotic-wavefunction}), where $C{=}\frac12{+}\Gamma$ here. The remaining terms are organized in terms of the small  expansion parameter ${\tilde g}{=}xE^\frac12/\hbar$. The diagrams associated to these expansions are all topologies with a single boundary (its on the probe FZZT brane),   some number~$h$ of handles, and some number of (pairs of)  points representing  $\Gamma$ units of R--R flux insertion. Each such element introduces a factor of   ${\tilde g}^2$.  So for the case (1,2) here ({\it i.e.}, $\Gamma{=}0$) the diagrams are purely a genus expansion weighted with ${\tilde g}^{2h-2+1}$.  For half--integer $\Gamma$,  an interpretation in terms of non--orientable surfaces seems more natural (given that the $(0,2)$ and (2,2)  ({\it i.e.} $\Gamma{=}{\pm}\frac12$) JT supergravities are  themselves  time--reversal symmetric).

The natural topological counting parameter $g$ for the expansion  in this regime, is the inverse of what is more familiar from the high energy regime. After building the partition function, an integration over $x$ and the Laplace transform will exchange $x$ for $\mu$, and $E$ for $1/\beta$, and the natural expansion parameter will be ${\hat g}{=}\mu/(\hbar\beta^\frac12).$  In JT supergravity it will simply count how many times an FZZT brane $(\psi)$ is included in an amplitude (basically a pair for every application of the operator ${\cal P}$).

\section{Low temperature}
\label{sec:low-temperature}
This section's goal  is an understanding of the thermal properties of the neighbourhood of the ground state of the JT supergravity theory, particularly the novel (1,2) case where there is a finite non--zero energy density at $E{=}0$. It will transpire that some insights will be gained that will be applicable to ordinary JT gravity too. The intuition developed in section~\ref{sec:sff-and-wormholes} and ref.~\cite{Johnson:2020exp} about the interplay of the role of wormholes and non--perturbative physics will pay off considerably here.

The thermal physics can in principle all be computed from the free energy: 
\be
F=-\frac{1}{\beta}\langle \ln Z(\beta) \rangle\ .
\ee
To be clear, this is the ensemble average of the logarithm of the partition function, as opposed to the logarithm of the average of the partition function. This ensemble average can in principle be done by using the replica method:
\be
\langle \ln Z(\beta)\rangle = \lim_{n\to0}\left(\frac{\langle Z(\beta)^n\rangle - 1}{n}\right) \ .
\ee
While this is easy to state, it is a hard computation to do and interpret correctly using traditional gravity techniques, as discussed recently in this JT context in ref.~\cite{Engelhardt:2020qpv}. However, as will now be shown,  some progress can be made using minimal string technology.

\subsection{Preparing for Replicas}
The matrix model (minimal string) techniques of section~\ref{sec:toolbox} can in principle be used to compute $ \langle Z(\beta)^n\rangle$ for arbitrary temperatures, but there is an obstruction to progress in making sense of the replica definition above. In the full non--perturbative JT supergravity computations of ref.~\cite{Johnson:2020exp} and section~\ref{sec:JT-supergravity-results-12}, the wavefunctions are only known numerically, and so  the much needed $n$--dependence of the quantity is obscured. Even in the low energy and small~$\hbar$ limit of the previous section, where the wavefunctions can be written as Bessel functions (see equation~(\ref{eq:bessel-wavefunction})), the nested integrals defining $ \langle Z(\beta)^n\rangle$ are hard to massage into a form where the leading $n$ dependence becomes apparent.  Nevertheless, some significant progress can be made if only interested in a limit.

\begin{figure}[h]
\centering
\includegraphics[width=0.25\textwidth]{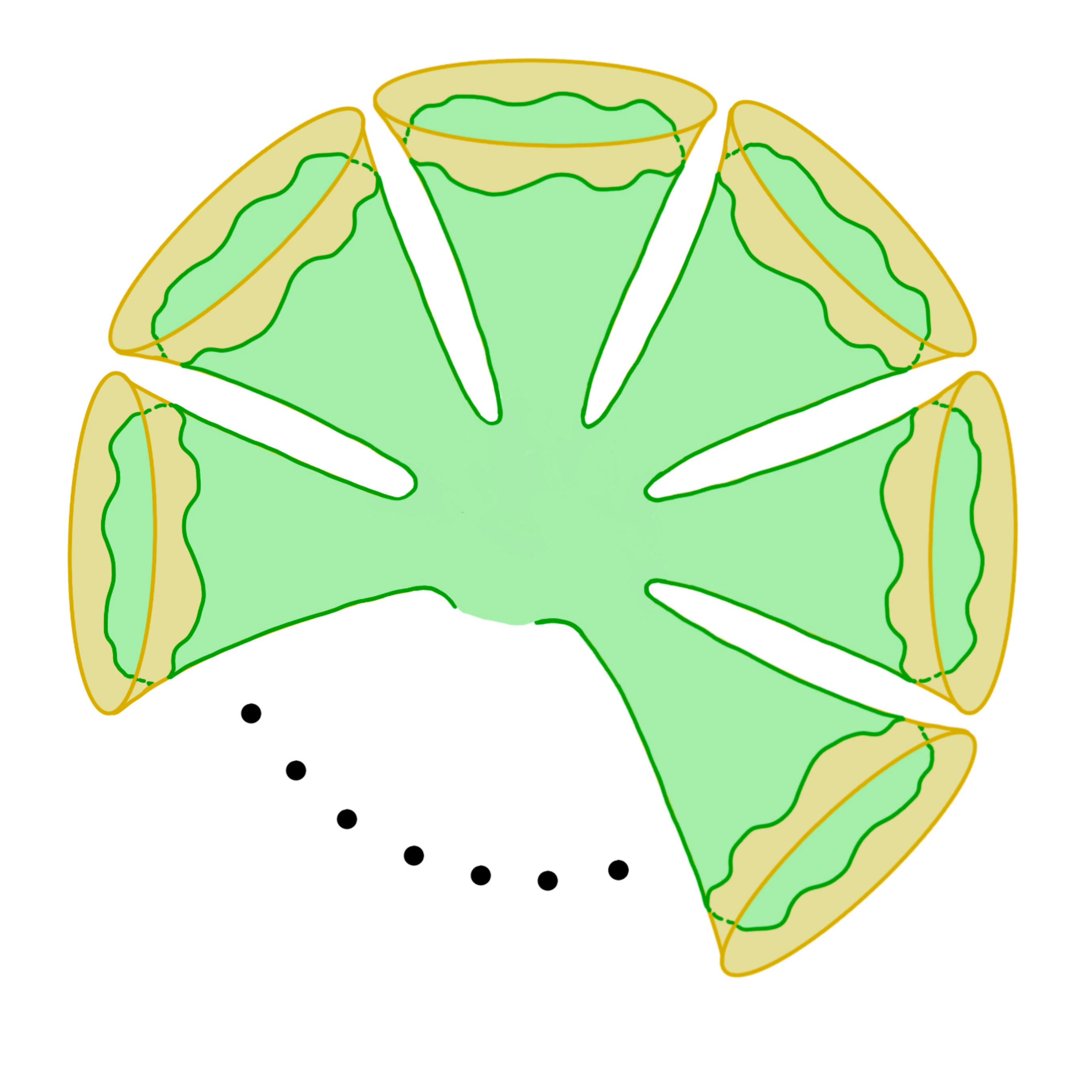}
\caption{\label{fig:n-point-wormhole} The (crystalline entity) $n$--point wormhole geometry, which dominates at low energy.}
\end{figure}

Since the goal is low temperature physics, the full wavefunctions are not needed. It is enough to expand them to leading order in a small energy expansion. It turns out that in this limit, all the integrals can be done! This will be enough to learn about the large $\beta$ (small~$T$) regime.  The next observation to be made is that, in a generalization of what was seen for the two--point function in section~\ref{sec:sff-and-wormholes}, the {\it fully connected} piece of the $n$--point correlator has the lowest power of $\beta$ in its denominator, so it dominates at ultra--low temperatures. See figure~\ref{fig:n-point-wormhole}. 
Therefore, contributions from pieces that mix connected and disconnected pieces can be ignored at leading order in the large $\beta$ expansion.

The general form of the connected correlator  for the product  $\prod_i^n Z(\beta_i)$ has an amusing form. It is implicit in ref.~\cite{Banks:1990df} (see also ref.~\cite{Moore:1991ir,Ginsparg:1993is}), but was put into a rather beautiful explicit form in ref~\cite{Okuyama:2018aij}:
\bea
\label{eq:connected-n-point}
&&\langle \prod_i^n Z(\beta_i)\rangle \\
&&\hskip0.3cm ={\rm Tr} \ln\left[ 1+\left\{ -1+\prod_{i=1}^n(1+z_ie^{-\beta_i{\cal H}})\right\}{\cal P}\right]_{{\cal O}(z_1z_2\cdots z_n)}\ ,\nonumber
\eea
where ${\cal P}$ is defined in equation~(\ref{eq:projector}). For the computation in hand,  all the $\beta_i$  are set to $\beta$. It is worth looking at the structure of some examples:
\bea
\label{eq:multi_Z}
\langle Z(\beta)\rangle &=&{\rm Tr}[ e^{-\beta{\cal H}}{\cal P}] \nonumber\\
\langle Z(\beta)^2\rangle &=&{\rm Tr}[ e^{-2\beta{\cal H}}{\cal P}]-{\rm Tr}[ e^{-\beta{\cal H}}{\cal P}e^{-\beta{\cal H}}{\cal P}]\nonumber\\
\langle Z(\beta)^3\rangle &=&{\rm Tr}[ e^{-3\beta{\cal H}}{\cal P}]-3{\rm Tr}[ e^{-\beta{\cal H}}{\cal P}e^{-2\beta{\cal H}}{\cal P}]\nonumber\\
&&\hskip1cm +2{\rm Tr}[ e^{-\beta{\cal H}}{\cal P}e^{-\beta{\cal H}}{\cal P}e^{-\beta{\cal H}}{\cal P}] \nonumber\\
\langle Z(\beta)^4\rangle  &=&{\rm Tr}[ e^{-4\beta{\cal H}}{\cal P}]-4{\rm Tr}[ e^{-\beta{\cal H}}{\cal P}e^{-3\beta{\cal H}}{\cal P}]\nonumber\\
&&\hskip1.8cm -3{\rm Tr}[ e^{-2\beta{\cal H}}{\cal P}e^{-2\beta{\cal H}}{\cal P}]\nonumber\\
&&\hskip0.7cm +12{\rm Tr}[ e^{-2\beta{\cal H}}{\cal P}e^{-\beta{\cal H}}{\cal P}e^{-\beta{\cal H}}{\cal P}] \nonumber\\
&&\hskip-0.05cm -6{\rm Tr}[ e^{-\beta{\cal H}}{\cal P}e^{-\beta{\cal H}}{\cal P}e^{-\beta{\cal H}}{\cal P}e^{-\beta{\cal H}}{\cal P}] \ .
\eea
The next step is to turn these traces into integrals over the explicit forms of the wavefunctions. This goes as before for the one-- and two-- point case described in sections~\ref{sec:partition-function} and~\ref{sec:sff-and-wormholes}. First, inserting the projector ${\cal P}$ results in, {\it e.g.:}
\bea
&&{\rm Tr} [e^{-m_1\beta{\cal H}}{\cal P} e^{-m_2\beta{\cal H}} {\cal P}\cdots e^{-m_M\beta{\cal H}} {\cal P} ] = \\
&&\hskip0.1cm\int dx_1 \int dx_2 \cdots \int dx_M\times\nonumber\\
&& \langle x_1| e^{-m_1\beta{\cal H}}| x_2\rangle \langle x_2| e^{-m_1\beta{\cal H}}|x_3\rangle  \cdots \langle x_{M}|e^{-m_1\beta{\cal H}}|x_1\rangle  \nonumber
\eea

Inserting a complete set of wavefunctions {\it via}: $1{\equiv}\int d\psi |\psi\rangle \langle \psi |$ after every exponential and writing $\langle x|\psi(E)\rangle {\equiv}\psi(x,E)$  results in a family of nested integrals (essentially Laplace transforms for inverse temperatures~$m_i\beta$) over the energies, with integrands built out of products of objects of the form
\be
\label{eq:integration}
\rho(E^\prime, E)\equiv \int_0^{\bar\mu}\!\!\!\psi(E^\prime,x)\psi^*(E,x) dx\ .
\ee
Notice that the upper limit on $x$ is denoted ${\bar\mu}$, the value of which will be determined later. Intuitively, ${\bar\mu}$ should be expected to depend on $n$ in some way. To understand why, it is useful to recall  the matrix model origin of the coordinate $x$. Before taking the double scaling limit, there is a discrete label or index, $i$ which runs from~$1$ to the size of the matrix, $N$. In taking $N{\to}\infty$, the label~$i$ is exchanged for a continuous label/coordinate  $X{=}i/N$  that runs from $0$ to $1$. The ends of the (unscaled) distribution of matrix energies are located at $X{=}1$. The double scaling limit zooms into the endpoint, and the coordinate~$x$ arises as $X{=}1{-}x\delta^2$, where $\delta{\to}0$ as $N{\to}\infty$. (The parameter $\hbar$ arises as the scaled $1/N$ topological expansion parameter {\it via} $1/N{=}\hbar \delta^q$, where $q$ is a positive number.) The point here is that the scale of~$x$ is tied to the original size of the matrix $N$. In doing the replica method here to perform the averaging, the formalism is being used to compute for $n$ copies of the large $N$ system, so to compare to physical properties already computed by the matrix model for a single copy of the formalism, it makes sense that $x$ (and hence the integration limit ${\bar\mu}$) will need to be scaled with~$n$ in some fashion. This will become apparent below.

\subsection{Free Energy from Replicas}
\label{sec:replica-scaling-limit}
The next step is to simply perform all the many integrals that need to be done. As mentioned earlier, the integrals are able to be performed if the wavefunctions are first expanded in small~$E$. Inserting the   low energy expanded wavefunction~(\ref{eq:wave-function-small-E-12}) gives for the $(1,2)$ case:

\bea
\label{eq:multi-Z-12}
\langle Z(\beta)\rangle &=&\frac{1}{4}\frac{{\bar\mu}^2}{\hbar^2}\frac{1}{\beta}-\frac{1}{16} \frac{{\bar\mu}^4}{\hbar^4}\frac{1}{\beta^2} 
+\cdots    \nonumber\\
\langle Z(\beta)^2\rangle &=&\frac{1}{8}\frac{{\bar\mu}^2}{\hbar^2}\frac{1}{\beta}-\frac{5}{64} \frac{{\bar\mu}^4}{\hbar^4}\frac{1}{\beta^2} 
+\cdots\nonumber\\
\langle Z(\beta)^3\rangle &=&\frac{1}{12}\frac{{\bar\mu}^2}{\hbar^2}\frac{1}{\beta}-\frac{29}{288} \frac{{\bar\mu}^4}{\hbar^4}\frac{1}{\beta^2} 
+\cdots\nonumber\\
\langle Z(\beta)^4\rangle  &=&\frac{1}{16}\frac{{\bar\mu}^2}{\hbar^2}\frac{1}{\beta}-\frac{103}{768} \frac{{\bar\mu}^4}{\hbar^4}\frac{1}{\beta^2} 
+\cdots
\eea
where recall that the range on the projector $x$--integral is $0{\leq}x{\leq}{\bar\mu}$, where ${\bar\mu}$ will be discussed below.

Similarly, using equation~(\ref{eq:wave-function-small-E-22}) for the $(2,2)$ case the results are:
\bea
\label{eq:multi-Z-22}
\langle Z(\beta)\rangle &=&\frac{1}{6}\frac{1}{\sqrt{\pi}}\frac{{\bar\mu}^3}{\hbar^3}\frac{1}{\beta^\frac32}-\frac{1}{20}\frac{1}{\sqrt{\pi}} \frac{{\bar\mu}^5}{\hbar^5}\frac{1}{\beta^\frac52} +\cdots    \nonumber\\
\langle Z(\beta)^2\rangle &=&\frac{1}{24}\frac{1}{\sqrt{\pi}}\frac{{\bar\mu}^3}{\hbar^3}\frac{1}{\beta^\frac32}-\frac{1}{160}\frac{1}{\sqrt{\pi}} \frac{{\bar\mu}^5}{\hbar^5}\frac{1}{\beta^\frac52}+\cdots\nonumber\\
\langle Z(\beta)^3\rangle &=&\frac{1}{54}\frac{1}{\sqrt{\pi}}\frac{{\bar\mu}^3}{\hbar^3}\frac{1}{\beta^\frac32}-\frac{1}{540}\frac{1}{\sqrt{\pi}} \frac{{\bar\mu}^5}{\hbar^5}\frac{1}{\beta^\frac52}+\cdots\nonumber\\
\langle Z(\beta)^4\rangle  &=&\frac{1}{96}\frac{1}{\sqrt{\pi}}\frac{{\bar\mu}^3}{\hbar^3}\frac{1}{\beta^\frac32}-\frac{1}{1280}\frac{1}{\sqrt{\pi}} \frac{{\bar\mu}^5}{\hbar^5}\frac{1}{\beta^\frac52}+\cdots
\eea
and finally,  using equation~(\ref{eq:wave-function-small-E-02}) for the $(0,2)$ case:
\bea
\label{eq:multi-Z-02}
\langle Z(\beta)\rangle &=&\phantom{\frac{1}{1}}\frac{1}{\sqrt{\pi}}\frac{{\bar\mu}}{\hbar}\frac{1}{\beta^\frac12}-\frac{1}{6}\frac{1}{\sqrt{\pi}} \frac{{\bar\mu}^3}{\hbar^3}\frac{1}{\beta^\frac32} +\cdots    \nonumber\\
\langle Z(\beta)^2\rangle &=&\frac{1}{2}\frac{\sqrt{2}}{\sqrt{\pi}}\frac{{\bar\mu}}{\hbar}\frac{1}{\beta^\frac12}-\frac{1}{\pi} \frac{{\bar\mu}^2}{\hbar^2}\frac{1}{\beta} +\cdots    \nonumber\\
\langle Z(\beta)^3\rangle &=&\frac{1}{3}\frac{\sqrt{3}}{\sqrt{\pi}}\frac{{\bar\mu}}{\hbar}\frac{1}{\beta^\frac12}-\frac{3}{2}\frac{1}{\pi} \frac{{\bar\mu}^2}{\hbar^2}\frac{1}{\beta} +\cdots    \nonumber\\
\langle Z(\beta)^4\rangle &=&\frac{1}{4}\frac{\sqrt{4}}{\sqrt{\pi}}\frac{{\bar\mu}}{\hbar}\frac{1}{\beta^\frac12}-\frac{1}{\pi} \frac{{\bar\mu}^2}{\hbar^2}\frac{1}{\beta} +\cdots  
\eea
In each case, notice that, as promised at the end of subsection~\ref{sec:low-energy-tail-tip}, the  expansion parameter is ${\hat g}{=}{\bar\mu}/\hbar\beta^\frac12$. It is very natural from the underlying minimal string perspective, where the physics  is simply closed strings. A pair of probe D--branes $(\psi)$ is inserted with every action of ${\cal P}$ in forming the correlator of a string of $Z(\beta)$s, giving a factor of ${\hat g}^2$ each time. There is  an overall extra ${\hat g}^{2\Gamma}$ (analogous to the extra flux insertion, or pair of cross--caps), and this gives the organization of the above expansions.

For the $(1,2)$ case, the $n$ dependence of the leading term is evidently such that
\be
\langle Z(\beta)^n\rangle =\frac{1}{4n}\frac{{\bar\mu}^2}{\hbar^2}\frac{1}{\beta}+\cdots
\ee
(The subleading pieces will be discussed in a moment.) Looking ahead to the replica  method, this will give
\be
\langle \ln Z(\beta)\rangle = \lim_{n\to0} \left(\frac{1}{4n^2}\frac{{\bar\mu}^2}{\hbar^2}\frac{1}{\beta}  - \frac{1}{n}\right) +\cdots\ ,
\ee
where again, the dots contain subleading contributions from both connected and disconnected contributions. This expression appears to be badly divergent. However, the expected $n$ dependence of ${\bar\mu}$ (anticipated  below equation~(\ref{eq:integration})) has not yet been  yet determined.  Requiring a finite result is a natural way of determining its behaviour. The most obvious choice is that it should scale like $n$, {\it i.e.} ${\bar\mu}{=}n\mu$, yielding a finite contribution from the first term.  

The  last  term, also divergent,  cannot be dealt with in the same way.  It is not uncommon to encounter (non--universal) divergent pieces in taking the double scaling limit, but there could be an admixture of  important physics obscured by it too. Since it is $\beta$--independent, it will not affect the discussion (below) of the leading behaviour of $F$ with temperature. An example of  another divergent piece that should be handled carefully is as follows. Ultimately  a $\beta$--independent piece will yield a constant contribution to the entropy. But this is already understood to be the extremal entropy $S_0{=}{-}\ln\hbar$. It can be naturally  included here. To understand how it fits, it is again useful to recall the double--scaled origins of this quantity.  The~$N$ states of the system should produce an entropic $\ln N$ contribution to the average $\langle \ln Z\rangle$. On the other hand, the double--scaling limit involves writing $1/N{=}\hbar\delta^q$ ($q$ some positive number) as $\delta{\to}0$. So $\ln N{=}{-}\ln\hbar{-}\ln \delta^q=S_0-\ln\delta^q$, and the latter term is of the same (but opposite sign) character as the divergence being seen in the replica result above. Put differently, a fixed additive ambiguity can be absorbed into  $S_0$.  

Finally, then,  the leading result for the free energy is:
\be
\label{eq:free-energy-result-12}
F = -\frac{1}{\beta}\langle \ln Z(\beta)\rangle =-\frac{S_0}{\beta}-\frac{1}{4}\frac{\mu^2}{\hbar^2}\frac{1}{\beta^2}  \ ,
\ee
a satisfying outcome. Note that a subtle  feature that worked well for this method is the fact that the expansion of the wavefunctions in low $E$ meant that the range of $x$ should be small for the expressions to remain valid. So having ${\bar\mu}$ shrink as the replica limit $n{\to}0$ is performed is consistent with the approximations made. 

Before moving on, a glance at the first level of subleading terms in the expansions in equation~(\ref{eq:multi-Z-12}) (still from the connected correlator) shows an interesting pattern. The terms seem to be  actually $1/16 n$ times a number that is the coefficient of the $n$th term in an expansion of $\ln(1{-}m)/(1{-}2m)$ (this is not proven, but the first four terms match). But this can't be the whole story, since there will also be terms at these orders coming from diagrams that mix connected and disconnected pieces. It would be interesting to see what the overall $n$--dependence looks like. 
 This will  be left for future work.

The leading order free energy~(\ref{eq:free-energy-result-12}) produces an entropy that is linear in its dependence on $T$:
\be
S(\beta) = \left( 1-\beta\frac{\partial}{\partial\beta}\right)\langle\ln Z(\beta) \rangle = S_0+\frac{1}{2} \frac{\mu^2}{\hbar^2} T\ ,
\ee
and so a linear specific heat also:
\be
C=T\frac{\partial S}{\partial T} = \frac{1}{2} \frac{\mu^2}{\hbar^2} T\ .
\ee
Notably, the entropy is manifestly positive all the way down to where it becomes the extremal value $S_0$ at $T{=}0$. This is in contrast to the logarithmically divergent negative entropy that  comes from using $\ln\langle Z(\beta)\rangle$ as the free energy, as discussed in the introduction.

This result is also consistent with the earlier observed density of states at $E{=}0$, which was in this limit:
\be
\rho(0)=\frac14\frac{\mu^2}{\hbar^2}\ ,
\ee
showing the same dependence on the parameters. This finite energy density enters all the thermodynamic quantities in the neighbourhood of $T{=}0$. For example, the energy $U{=}{-}\partial_\beta\langle\ln Z(\beta)\rangle$ is:
\be
U=\frac14\frac{\mu^2}{\hbar^2} T^2\ .
\ee
See the Introduction for some discussion (below equation~(\ref{eq:replica-results})) of the interpretation of this result.

 This apparent success for the toy (1,2) case should  be contrasted with its (2,2) and (0,2) cousins. For the (2,2) case, the $n$ dependence of the leading term is 
\be
\langle Z(\beta)^n\rangle =\frac{\sqrt{n}}{6n^2}\frac{{\bar\mu}^3}{\hbar^3}\frac{1}{\sqrt{\pi}}\frac{1}{\beta^\frac32}+\cdots
\ee
The same choice as before, ${\bar\mu}{=}n\mu$, will lead to a vanishing (at this order in $\beta$) of the free energy in the $n{\to}0$ limit, because of the extra $n\sqrt{n}$ factor. This would be consistent with there being no finite $\rho(E)$ at $E{=}0$. Perhaps higher orders in $\beta$ will yield something non--vanishing in the limit.  
For the (0,2) case: 
\be
\langle Z(\beta)^n\rangle =\frac{\sqrt{n}}{n}\frac{{\bar\mu}}{\hbar}\frac{1}{\sqrt{\pi}}\frac{1}{\beta^\frac12}+\cdots
\ee
This time, choosing the same ansatz ${\bar\mu}{=}n\mu$ would again give a vanishing free energy in the $n{\to}0$ limit, since there is an uncanceled $\sqrt{n}$. Again, working at higher orders in temperature should yield something non--vanishing.\footnote{Alternatively, in an attempt to get a non-zero free energy for these models, a different scaling with $n$ of $\bar\mu$ can be chosen to make the $\beta$ dependent piece finite, resulting in an $F$ that scales as $T^{3/2}$ for (0,2) or  $T^{5/2}$ for (2,2). This  seems less well--motivated, not the least because this amounts to changing the double-scaled matrix model definition for different choices of~$\Gamma$.}

\subsection{Low Energy and Temperature for  JT Gravity}
\label{sec:JT-at-low-T-and-E}
The success seen in the last section for the  ``replica--scaling'' method encourages a search for other examples. The most obvious one is JT gravity itself. There are subtleties, because there are non--perturbative instabilities in the standard matrix model definition given in ref.~\cite{Saad:2019lba}, connected to the fact that there are non--perturbative contributions to the spectral density for $E{<}0$. A stable non--perturbative definition was proposed in ref.~\cite{Johnson:2019eik}, and examined further in ref.~\cite{Johnson:2020exp}. For that case, the  spectrum stops at $E{=}0$, and in fact the spectral density was again observed to be non--zero and finite there. 

If the  replica method presented here is robust, it should produce a similar low temperature thermodynamics to that seen for the (1,2) supergravity, since the properties derived were attributed to the presence of $\rho(0)$, the finite $E{=}0$ density. This is worth trying to get to work. A quick and instructive way to proceed is to use the observation that  in ref.~\cite{Johnson:2019eik}, once the dust had settled, the resulting spectral density for the low energy $k{=}1$ tail looked rather like that of the Airy case, but simply truncated at $E{=}0$ (see figure~4 there). This is  not exactly right of course: At low enough energies, the Airy functions handed over to Bessel functions, which  naturally cut off the  undesirable part of the spectrum\footnote{This all worked nicely because the underlying minimal models used were the $(4k,2)$ models also used to later build the supergravity models, but combined together in a way that yields the bosonic spectrum at higher energies. Simply put, this definition of JT gravity ``borrows'' the good low energy behaviour of JT supergravity.}. 

So an instructive  analysis of the low energy spectrum of JT gravity that avoids the non--perturbative bad behaviour is to treat the low energy $k{=}1$ tail as   being the usual Airy model, but truncate at $E{=}0$. The wavefunction for the Airy case is given in equation~(\ref{eq:airy-wavefunction})
and expanding about small $E$ and $x$ gives: 
\be
\psi(E,x)\simeq\frac{1}{3^\frac23\hbar^\frac23\Gamma(\frac23)}+\frac{3^\frac16\Gamma(\frac23)}{\hbar^\frac43\pi}(x+E)+\cdots\ ,
\ee
(here, $\Gamma$ denotes the $\Gamma$--function) and the next step is to insert it into the multi--point functions and compute the integrals. This time the projector~${\cal P} $ is different. Instead it is:
\be
{\cal P} = \int_{\bar x}^0 dx |x\rangle\langle x|\ ,
\ee
where ${\bar x}$ will not go to $-\infty$ as it usually does since the expansion of the wave--function is only valid for small $x$. This is appropriate since then the result will be sensitive only to the low energy physics at $E$ and $x$ small.

The leading order (in large $\beta$) result is  (neglecting anything higher order than linear in $\bar x$):
\be
\langle Z(\beta)^n\rangle =\frac{1}{9n}\frac{|{\bar x}|}{\beta}\frac{3^\frac23}{\hbar^\frac43\Gamma(\frac23)^2}+\cdots
\ee
and so  if the scaling  choice $|{\bar x}|{=}n$ is made (again, consistent with previous expectations), the free energy that results is (after including the $S_0$ term and treating the $1/n$ term as done above for $(1,2)$):
\be
F= -\frac{S_0}{\beta}- \frac{1}{3^{\frac43}\hbar^\frac43\Gamma(\frac23)^2}\frac{1}{\beta^2}\ ,
\ee
giving  again an entropy and specific heat that are linear in~$T$, and a quadratic energy $U$. Since the use of the Airy wavefunction here was somewhat heuristic, the pure numbers in the $1/\beta^2$ term should not be taken too seriously. However, the factors of $\hbar{=}e^{-S_0}$ are instructive---notice that  the combination $e^{2S_0/3}T$  naturally appears, matching elements of the replica wormhole discussion in ref.~\cite{Engelhardt:2020qpv}. This work  will be discussed in the next section. 

The actual non--perturbative definition of JT gravity presented in ref.~\cite{Johnson:2019eik} used the $(4k,2)$ minimal models in an appropriate combination, and so its low $T$ physcis  can be given the same treatment as the previous section for $\Gamma{=}0$. The difference is that its natural $\mu$ value is zero (there's no ``tip" of the tail), which would give a vanishing  free energy at this order. To make progress in this case (to explore the effects of the finite non--vanishing~$\rho(0)$ that it has), the small $\hbar$ limit would need to be relaxed somewhat, and wavefunctions built from solutions of Painlev\'e~II according to section~\ref{sec:Painleve-II} would need to be used instead. Perhaps  keeping some extra terms in the asymptotic expansion for $v(x)$ could yield expressions for $\psi(x,E)$ that  give analytic results for~$\langle Z(\beta)^n\rangle$.

\section{Conclusions}
\label{sec:conclusions}

A core result  is that the spectral density $\rho(E)$ of (1,2) JT supergravity (and a non--perturbatively regularized proposal for ordinary JT gravity) has a finite non--vanishing piece at zero energy, $\rho(0)$, generated by non--perturbative quantum effects. This is particularly interesting, not the least because it could be a simple model of various kinds of disordered or frustrated  systems with either exactly this property, or close to it. The thermodynamics uncovered matches some of their characteristics~\cite{doi:10.1080/14786437208229210,PhysRevLett.39.41}. Such systems often have glassy behaviour. The ability to use quantum gravity to study the non--trivial low temperature dynamics of the kind found here seems like a valuable opportunity for modelling aspects of quantum disordered phases. It is complementary to the approach of refs.~\cite{Anninos:2016szt,Anninos:2013mfa,Anninos:2012gk,Anninos:2011vn}.

The results obtained here for the free energy $F(T)$, summarized in equation~(\ref{eq:replica-results}) (see also the discussion below it), apply to a very simplified toy model of the tail of the spectral density for which the leading low energy/temperature behaviour is simply $\rho(E)=\rho(0)+\cdots$ with $Z(\beta){=}\rho(0)/\beta+\cdots$, a non--perturbative feature shared by $(1,2)$ JT supergravity and also ordinary JT gravity\footnote{At the very least in the non-perturbative definition supplied in refs.~\cite{Johnson:2019eik,Johnson:2020exp}.}. In this limit, the wavefunctions were simple enough to compute the results exactly. The single energy scale that emerges, $1/\rho(0)$, then naturally controls the leading low energy dynamics. The other two models considered, (2,2) and (0,2) JT supergravity, do not have a finite non-zero $\rho(0)$ in this limit, and so this goes some way to understanding why a non-zero $F$ did not emerge for them using these methods. Working with the complete Bessel models of these systems (not just the leading part) will yield other non-perturbative scales that will likely modify the results of this approach. This would be interesting to pursue.

While the method of scaling $\bar\mu$ with $n$ to get a finite contribution for $F$'s $\beta$--dependence obtained sensible results, it is understandably a little unsettling that the  $\beta$--independent divergent $1/n$ piece was able to be neglected without adverse consequences. It does  mean that important finite $\beta$-independent  contributions to $F$ can't be captured without better understanding of how to handle such terms. It is therefore very worthwhile to try to better understand this method, and place it on a firmer footing.\footnote{As mentioned in the note added in the Introduction, subsequent work~\cite{Johnson:2021rsh} using different computational methods has confirmed the results obtained in this paper, suggesting that this replica-scaling method is robust. Moreover, the low energy non-perturbative scales present in the complete Bessel models and also in full models of JT gravity and JT supergravity  can be captured, in a manner that follows naturally from what was seen in this paper.}

It is worth noting that this successful  combined replica and double--scaling  method  had a surprising simplicity (at least to leading order in the large~$\beta$ expansion). In particular, the $n$ dependence of the $\langle Z(\beta)^n\rangle$ quantity is of a rather simple and unambiguous form. It followed from the fact that in the large~$\beta$ limit, the dominant contribution to the connected correlator is the  part that treats the~$n$ boundaries symmetrically. This translated nicely into the fact that there was no need for ``replica symmetry breaking''~\cite{Parisi:1979mn,Parisi:1979ad} arising from a need to make sense of the availability of different analytic continuations of the~$n$ dependence, as happens in {\it e.g.}, the classic spin glass models of refs.~\cite{Edwards_1975,PhysRevLett.35.1792}.  This  suggests that while there are some shared key features with simple models of spin glasses (linear $S$ and~$C$, non--zero finite $\rho(0)$, {\it etc.}), this part of the phase diagram ({\it i.e.,} linearly emerging from $T{=}0$) is somewhat different. This does not exclude similarities emerging at nearby regions of the phase diagram (see below). 

 The possible meaning of this paper's results for models such as SYK and generalizations thereof would be interesting to establish. It would also connect to recent discussions  ({\it e.g.} in refs.~\cite{Gur-Ari:2018okm,Arefeva:2018vfp,Caracciolo:2018rwg,Baldwin:2019dki} about the difficulty of  finding spin glass phases in  SYK and SYK--like  models. Note that some of the arguments~\cite{Gur-Ari:2018okm} center around the fact that there is random matrix behaviour lurking, which seems to be incompatible with the phenomenology of glassy behaviour (due to eigenvalue repulsion). Note however that non--perturbative effects have been explicitly shown here and in ref.~\cite{Johnson:2020exp} (using random matrix models, ironically) to yield a breakdown of eigenvalue repulsion at $E{=}0$, so there seems to be an opportunity.

Meanwhile,  ref.~\cite{Engelhardt:2020qpv} presented a recent discussion of replica wormholes and replica symmetry breaking in the context of JT gravity, where the gravity computations are done explicitly. They interpret some of their findings as a sign of replica symmetry breaking.  As mentioned above, by building the quantum gravity theory path integral using matrix model methods, the issues they had seem to have been avoided. It is not at present clear whether this success is intrinsic to the matrix model approach's known ability to sidestep  technical difficulties and get to the answers directly, or whether it is because of better control of non--perturbative aspects of gravity.  It may well be a mixture of the two, although the analysis of this paper shows that significant  non--perturbative effects in the low temperature regime play a dominant role in the physics, and so it is difficult to compare to their replica computation in order to see where the point of departure may be. This is certainly worth further exploration. 

It is also possible that at the next non--trivial order in $1/\beta$, the replica computation with the matrix model approach also gets more complicated. This is because there will be terms of two kinds appearing. Mixtures of connected and disconnected diagrams, as well  as connected diagrams where (looking at the structure of equations~(\ref{eq:multi_Z}) for example) different subsets of the $n$--boundaries get different emphasis. So the $n$--dependence is likely to be less trivial, and hence subtleties may occur in the analytic continuation in preparation for taking the $n{\to}0$ limit that may well amount to replica symmetry breaking, arising literally from wormhole sectors treating the $n$--boundaries unequally. This could give an opportunity for at least some of  the physics seen in spin glasses to appear at higher $T$.  It is interesting that the leading order effects studied in this paper managed to avoid this.

For future work, it could  be worthwhile to perform the replica method directly in the matrix model {\it before} double scaling, afterward taking the double scaling limit to return to the continuum theory. The method here was guided somewhat by knowledge of the un--double--scaled theory, but was a hybrid of both sides of the limit. In addition to possibly locating clues as to how the replica symmetry breaking issues mentioned above were avoided, such an approach could also yield  in a nicely organized  toolbox  for obtaining  higher order $T$  and $\hbar$ corrections.

The linearity of the entropy and specific heat seen at low $T$ in this paper is of a different slope than that of the high temperature phase. So there is a possible phase transition or interesting crossover in the intermediate regime to be understood.  It could be connected to the results of ref.~\cite{Engelhardt:2020qpv} and/or the subleading physics discussed two paragraphs above. Again, the possibility of more lessons from spin glasses may well be alive. 

Replica methods of various kinds are powerful tools to bring to bear in  understanding various aspects of quantum gravity. Now that double--scaled matrix models have been (again) rebooted in the service of understanding quantum gravity, it is worthwhile further exploring replica computations with them as a  possible probe of some   observables~\footnote{There is  forthcoming work of this nature  by  Akash Goel and Herman Verlinde.}.

  \begin{acknowledgments}
CVJ  thanks   Netta Engelhardt, Robie Hennigar, Juan Maldacena, Alex Maloney, Felipe Rosso, Andy Svesko, Herman Verlinde and Edward Witten for  helpful remarks and discussions, the  US Department of Energy for support under grant  \protect{DE-SC} 0011687, and, especially during the pandemic,  Amelia for her support and patience.    
\end{acknowledgments}

 \appendix
 \section{Disconnected and Connected Pieces\\ of the (1,2) JT Supergravity Spectral Form Factor}
 \label{sec:extra-stuff}
\begin{figure}[h]
\centering
\includegraphics[width=0.48\textwidth]{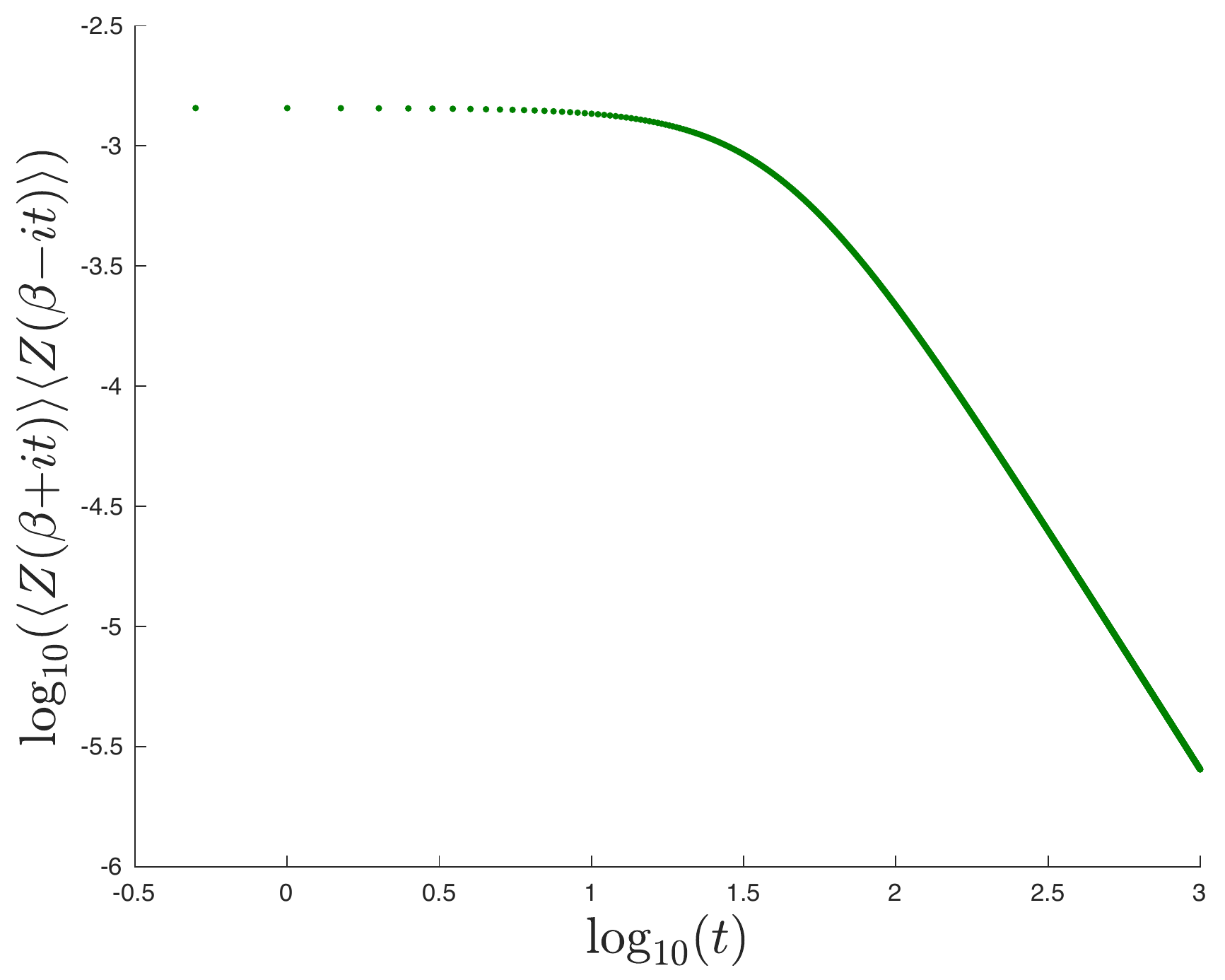}
\caption{\label{fig:disconnected-sff-SJT} The disconnected part of the  (1,2) JT supergravity  spectral form factor  {\it vs.} $t$, at $\beta{=}50$ and $\hbar{=}1$.}
\end{figure}

\begin{figure}[h]
\centering
\includegraphics[width=0.48\textwidth]{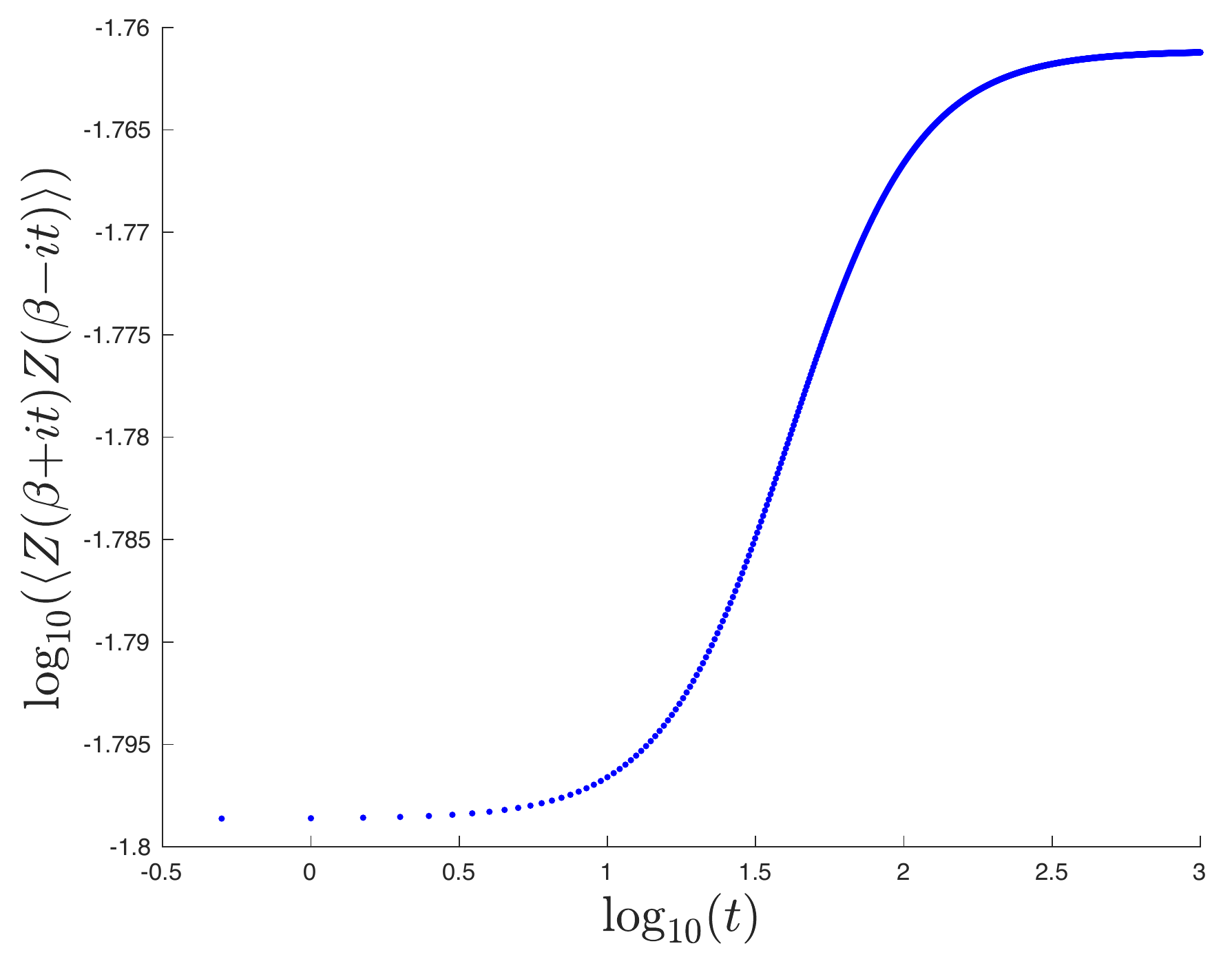}
\caption{\label{fig:connected-sff-SJT} The connected part of the (1,2) JT supergravity  spectral form factor  {\it vs.} $t$, at $\beta{=}50$ and $\hbar{=}1$.}
\end{figure}

 So as to not disrupt the flow of the main body of the paper, and since they were not immediately needed, the individual parts of the spectral form factor for the (1,2) JT supergravity are presented here. See subsection~\ref{sec:spectral-form-factor} for discussion, ref.~\cite{Johnson:2020exp} for the results for the (2,2) and (0,2) models,  and figure~\ref{fig:combined-sff-SJT} for the behaviour of their sum. The disconnected piece is given in figure~\ref{fig:disconnected-sff-SJT}, and the connected piece in figure~\ref{fig:connected-sff-SJT}.

\vfill\eject
\newpage

\bibliographystyle{apsrev4-1}
\bibliography{LE_super_JT_gravity,NP_super_JT_gravity}

\end{document}